\documentstyle[aps,eqsecnum,amssymb,prb]{revtex}

\input epsf

\newcommand{\bz}{\bbox{\zeta}}			

\newcommand{\rvec}{{\bf r}}
\newcommand{\rrvec}{{\bf r'}}			
\newcommand{\kvec}{{\bf k}}

\newcommand{\Kinv}[1]{{\bf K}^{-1}_{#1}}	
\newcommand{\Gind}[1]{{\bf G}_{#1}}		
\newcommand{\calGind}[1]{{\bbox{\cal G}}_{#1}}	
\newcommand{\Rind}[1]{{\bf R}_{#1}}		
\newcommand{\Sind}[1]{{\bf S}_{#1}}		
\newcommand{\Tind}[1]{{\bf T}_{#1}}		
\newcommand{\sigind}[1]{\bbox{\sigma}_{#1}}	
\newcommand{\rbarsub}[1]{\right|_{#1}}
\newcommand{\unitkk}{{\openone - \hat{\kvec}\hat{\kvec}}} 

\newcommand{\beq}{\begin{equation}}		
\newcommand{\eeq}{\end{equation}}
\newcommand{\beqar}{\begin{eqnarray}}
\newcommand{\eeqar}{\end{eqnarray}}
\newcommand{\la}{\left<}		
\newcommand{\ra}{\right>}		
\newcommand{\kk}{{\hat{\kvec}\hat{\kvec}}}

\newcommand{\Ginv}[1]{{\bf G}^{-1}_{#1}}

\newcommand{\unit}[1]{{\openone - \hat{\bf #1} \hat{\bf #1}}}

\newcommand{\rrrvec}{{{\bf r}''}}
\newcommand{\subs}[2]{{\bf #1}_{#2}}

\begin{document}
\draft

\title{Dynamics of Diblock Copolymers in Dilute Solutions}
\author{Radu P. Mondescu and M. Muthukumar}
\address{Department of Physics and Astronomy, and Polymer Science \& 
Engineering Department and Materials Research Science and Engineering Center \\
 University of Massachusetts, Amherst, MA 01003}
\maketitle
\date{\today}
\begin{abstract}
In the present work we consider the dynamics of freely translating and rotating
diblock (A--B), Gaussian copolymers, in dilute solutions.  Using the previously
developed multiple scattering technique \cite{FM78,FM278,MF79}---applied to the
study of suspensions of spheres and polymers---we have computed the diffusion
and the friction coefficients $D_{\text{AB}}$ and $\bz_{\text{AB}}$,
respectively, and the change $\delta\eta_{\text{AB}}$ in the viscosity of the
solution as functions of ${x = {N_A\over N}}$ and ${t = {l_B\over l_A}}$, where
$N_A$, $N$ are the number of segments of the A block and of the whole
copolymer, respectively, and $l_A$, $l_B$ are the Kuhn lengths of the A and B
blocks.  Specific regimes that maximize the difference between the diffusion
constants of copolymers with distinct $t$ values, which lead to
increasing the efficiency of separation processes, have been identified. 
\end{abstract}
\pacs{}

\section{Introduction}

Currently, due to both the practical significance and the theoretical
challenge, there is a sustained interest in the study of transport properties
of copolymers in solution, mainly in relation to electrophoresis and associated
phenomena. Basically, one wishes to {\sf tag} the macromolecules of interest
with a neutral chain to make the friction coefficient of the resulting object
molecular weight dependent for all measurement conditions, in order to increase
the efficiency of the separation process.\cite{ELECTRO} As there are
experimental regimes for which the molecular weight dependence is lost and the
separation is very weak if not impossible (for a recent calculation for
polyelectrolytes see Muthukumar\cite{MELECTRO}), a broader understanding of the
dynamics of composite objects in solution becomes critical.

To gain insight in the dynamical behavior of an arbitrary heterogeneous polymer
in solution, we will address a simpler, nevertheless illuminating problem:
consider two linear, Gaussian polymers A and B, with lengths/number of segments
$L_A$/$N_A$ and $L_B$/$N_B$, respectively, and Kuhn lengths $l_A$ and $l_B$,
joined to form a composite chain (A--B) of total length $L = L_A + L_B$ and
total number of monomers $N = N_A + N_B$. In this paper we calculate the
diffusion and the friction coefficients of the diblock copolymer A--B and the
change in the viscosity of the solution due to the copolymer in dilute
solutions, as functions of the dimensionless variables $t = {l_B\over l_A}$ and
$x = {N_A\over N}$.

Our calculations are based on the cluster expansion theory and on the effective
medium hypothesis.\cite{FM78,M81,MF82,M82,FM82} The main simplifying
assumption is the absence of excluded volume interaction. Other approximations
and features of our model are:
\begin{itemize}
\item[{\sl a\/})] no interaction between the polymer chains except the
hydrodynamic interaction is present.
\item[{\sl b\/})] the solvent is incompressible and the steady state limit is
considered.
\item[{\sl c\/})] the solution is assumed infinitely diluted (one chain limit).
\item[{\sl d\/})] the usual preaveraging approximation,\cite{YAM} where one
replaces the Oseen tensor ${\bf G}(\rvec,\rrvec)$ with its configurational
average $\la{\bf G}(\rvec,\rrvec)\ra$, is employed.
\item[{\sl e\/})] the hydrodynamic interaction is enforced by using stick
boundary condition.
\end{itemize}
\indent The model is not limited by the Gaussian nature of the two polymer
components, because the Kuhn lengths $l_A$ and $l_B$ could be {\sf effective}
quantities modeling chains with excluded volume and Coulombic interactions
(assertion valid in the hydrodynamic limit; at short distances the fractal
structure of the polymer becomes transparent). Also, even if the diblock
copolymer could be regarded as a particular example of a branched polymer,
extensively studied in the literature,\cite{YAM,FIX} the important distinction
in our work is that we allow for unequal Kuhn lengths for the two branches,
with interesting consequences in the calculations---the structure factor of the
copolymer must be computed now using a bivariate Gaussian probability
distribution function.  Finally, the present approach allows for obtaining
exact results, albeit numerical except some limits, and no approximation of
Kirkwood-Riseman type is necessary.

	Concerning the outline of the paper, we next develop
schematically---emphasizing the new elements---the underlying theory, then we
show explicitly the calculation of $D_{\text{AB}}$, $\bz_{\text{AB}}$ and
of $\delta\eta_{\text{AB}}$. We conclude with a discussion of the results.

\section{Calculation of the Steady-State Velocity Field: Infinite Dilution
Limit} 

The dynamical properties of the system can be deduced from the knowledge of the
average velocity field at any point in the solution. As this type of
computation is already very well documented,\cite{M81} we will show detailed
derivations only when necessary.  

Briefly, the main steps of our computations are: first we write the
velocity field in the solution in the presence of the copolymer; we apply the
boundary condition to eliminate the unknown friction forces between the
polymer and the fluid and then we average over the random position and the
configuration of the chain; next we construct the cluster expansion in terms of
the self-energy (Hartree-Fock diagrams) $\bbox{\Sigma}$ of the fluid and we
express it as a function of the known (measurable) physical parameters of the
system; finally, we use the self-energy to extract---both analytically and
numerically---the quantities of interest.

Assuming that the diblock copolymer A--B with total number of segments $N$ is
immersed in an ideal, incompressible fluid obeying the linearized, time
independent Navier-Stokes equation (with the solvent density equal to one), we
can write the equation followed by the velocity field ${\bf v}(\rvec)$
everywhere in the suspension as
\begin{equation}
- \eta_0 \triangle{\bf
v}(\rvec) + \nabla p(\rvec) = {\bf F}(\rvec) + \int_0^{L}\! ds\, \delta (\rvec
- \Rind{s}) \bbox{\sigma}_s \, , 
\label{navier}
\end{equation}
where ${\bf F}(\rvec)$ and $p(\rvec)$ are some external force producing the
flow and the pressure, respectively, $\eta_0$ is the kinematic shear viscosity
coefficient and $\sigind{s} = \bbox{\sigma}(s)$ is the force exerted by the
bead located at $\Rind{s}={\bf R}(s)$ upon the fluid, $s$ being the arclength
along the chain (for notational convenience we will subsequently write $s$ as a
subscript). The formal solution of Eq.~(\ref{navier})is readily given by the
integral representation (equivalent to the single layer solution of standard
boundary methods)
\begin{equation} 
{\bf v} (\rvec) = {\bf v}_0 (\rvec) + \int_{0}^{L}\! ds \, {\bf G}(\rvec -
\Rind{s})\cdot \sigind{s}\, ,
\label{velocity_sol}
\end{equation} 
in which ${\bf v}_0(\rvec) = \int\! d\rrvec \, {\bf G}(\rvec-\rrvec)\cdot {\bf
F}(\rrvec)$ is the externally imposed velocity field and ${\bf G}$ is the Oseen
tensor:
\begin{eqnarray} 
{\bf G}({\bf k}) = && \frac{\openone - \hat{\bf k}\hat{\bf k}}{\eta_0
k^2} \nonumber\\ 
{\bf G}({\bf r} - {\bf r'}) = {1\over 8 \pi \eta_0 |{\bf r} -
{\bf r'}|} && \left ( \openone + {({\bf r}-{\bf r'})({\bf r}-{\bf
r'})\over{|{\bf r}- {\bf r'}|^2}} \right) .
\label{Oseen}
\end{eqnarray} 
Here, $\openone$ is the unit tensor and $\hat{\kvec}$ is the unit vector
pointing in the $\kvec$ direction and the Fourier transform is given by
\begin{equation}
{\bf G}(\rvec) = \int\! {d\kvec\over (2 {\pi})^3} \, {\bf G}(\kvec) \exp(- i
\kvec\cdot\rvec). 
\label{Fourier_hydro}
\end{equation}

As mentioned, the dynamics of the copolymer chain and of the fluid are
coupled through the no-slip boundary condition 
\begin{equation}
\dot{\bf R}_s = \dot{\bf R}_0 + \bbox{\omega}\times\Sind{s} = {\bf v}(\Rind{s})
, \label{boundary}
\end{equation}
where the velocity of the center-of-mass $\dot{\bf R}_0$ is constant and we
further assume that the angular velocity $\bbox{\omega}$ is constant and equal
to the configuration averaged angular velocity of the chain. To this equation
we attach the physical constraints of force-free and torque-free motion
(neglecting the inertial terms) of the diblock chain:
\begin{eqnarray}
-{\bf f}\;(\text{total force}) & = & \int_0^L \!  ds\,\sigind{s} = 0
\nonumber\\ 
-{\bf M} \; (\text{total torque}) & = & \int_0^{L}\!  ds\,\Sind{s} \times
\sigind{s} = 0 .
\label{force_torque}
\eeqar
Above we used the relation $\Rind{s} = \Rind{0} + \Sind{s}$, where $\Sind{s}$
represents the position vector of the monomer at $s$ about the center-of-mass
($\Rind{0}$) of the copolymer.

Next one needs to eliminate the unknown forces $\sigind{s}$, a task
accomplished by using the boundary condition (\ref{boundary}) and defining the
single-chain inverse operator $\Ginv{}$ as
\begin{equation}
\int_0^L \! ds'\, {\bf G}^{-1}(\Rind{s},\Rind{s'})\cdot{\bf G}(\Rind{s'} -
\Rind{s''}) = \delta_{ss''} \openone .
\label{Oseen_inverse}
\end{equation}
We obtain then:
\begin{equation}
\sigind{s} = \int_0^{L}\! ds' \, {\bf G}^{-1}(\Rind{s},\Rind{s'})\cdot
({\dot{\bf R}}_0 + \bbox{\omega}\times\Sind{s'} - {\bf v}_0(\Rind{s'})),
\label{sigma_force}
\end{equation}
where the variables $s, s'$ should be read as the arclength arguments of ${\bf
R}(s)$, ${\bf R}(s')$. Inserting $\bbox{\sigma}$ back in (\ref{force_torque})
we can calculate the total force ${\bf f}$ and torque ${\bf M}$ acting upon the
polymer as
\begin{mathletters} 
\begin{eqnarray}
- {\bf f} & = & {\bf g}_T \cdot \dot{\bf R}_0 + {\bf g}_{TR}\cdot\bbox{\omega}
- \int\!\!\!\int_0^L\! ds\,ds'\, \Ginv{ss'}\cdot{\bf v}_0(s')
\label{total_force}\\  
- {\bf M} & = & {\bf g}_{RT}\cdot\dot{\bf R}_0 + {\bf g}_R\cdot\bbox{\omega} -
\int\!\!\!\int_0^L \! ds\,ds'\, \Sind{s}\times\Ginv{ss'}\cdot {\bf v}_0(s') ,
\end{eqnarray}
\end{mathletters}
where we manipulated $\Ginv{ss'}$ as a dyadic tensor. The newly introduced
tensors ${\bf g}_{T,R}$ provide, upon averaging over the distribution of the
segments $s$ and $s'$, the translational and rotational friction coefficients
of the copolymer:
\begin{mathletters}
\label{coeffs}
\begin{eqnarray}
\bz_{\text{AB}} & = & \la{\bf g}_T\ra = \la\int\!\!\!\int_0^L\! ds\,ds'\,
\Ginv{ss'}\ra \label{friction_coef}\\ 
\bz_{\text{AB}}^{\text{rot}} & = & \la {\bf g}_R \ra =  - \la
\int\!\!\!\int_0^L\!  ds\,ds'\, \Sind{s}\times\Ginv{ss'}\times
\Sind{s'}\ra\label{rot_coeff}.
\end{eqnarray}
\end{mathletters}
The remaining coefficients ${\bf g}_{TR,RT}$ (the cross
translational-rotational and rotational-translational terms) vanish upon
preaveraging, so will be further discarded. For future reference, their
expressions are:
\begin{mathletters}
\begin{eqnarray}
{\bf g}_{TR} & = & - \int\!\!\!\int_0^L \! ds\,ds'\,
\Ginv{ss'}\times\Sind{s'}\\ 
{\bf g}_{RT} & = & \int\!\!\!\int_0^L \! ds\,ds'\, \Sind{s}\times\Ginv{ss'}.
\end{eqnarray}
\end{mathletters}
Returning to the constraints (\ref{force_torque}), we now eliminate the
unknowns $\dot{\bf R}_0$ and $\bbox{\omega}$ using
\begin{eqnarray}
\dot{\bf R}_0 & = & {\bf g}_T^{-1}\cdot\int\!\!\!\int_0^{L} \! ds\,ds' \,
\Ginv{ss'} \cdot {\bf v}_0(s') \nonumber\\
\bbox{\omega} & = & {\bf g}_R^{-1} \cdot \int\!\!\!\int_0^{L}\! ds\, ds' \,
\Sind{s}\times \Ginv{ss'} \cdot {\bf v}_0(s') ,
\end{eqnarray}
where the inverse operators ${\bf g}_T^{-1}$ and ${\bf g}_R^{-1}$ are
defined by the relations ${\bf g}^{-1}_{T,R} \cdot {\bf g}_{T,R} =
\openone$. Inserting the values obtained for $\dot{\bf R}_0$ and
$\bbox{\omega}$ in Eq.~(\ref{sigma_force}) we find that
\begin{eqnarray}
\sigind{s} & = & - \int_0^L \! ds ' \, \left\{ \Ginv{ss'} -
\int\!\!\!\int_0^L\! dt \, dt' \, \Ginv{st} \cdot {\bf g}^{-1}_T \cdot \Ginv{t'
s '} \right. \nonumber\\ & & + \left. \int\!\!\!\int\! dt \, dt' \, \Ginv{st}
\times \Sind{t} \cdot {\bf g}^{-1}_R \cdot \Sind{t'} \times \Ginv{t's'}
\right\} \cdot{\bf v}_0(s')
\end{eqnarray}
By introducing this expression back in Eq.~(\ref{velocity_sol}) we get the
solution for the velocity field ${\bf v}(\rvec)$:
\begin{equation}
{\bf v}(\rvec) = {\bf v}_0(\rvec) + \int\!\!\!\int\!\!\!\int\! d\rrvec d\rrrvec
d\rrrvec ' \, {\bf G}(\rvec - \rrvec) \cdot {\bf T}(\rrvec,\rrrvec) \cdot {\bf
v}_0(\rrrvec) .
\label{sol_scatt}
\end{equation}
{\bf T}, the {\sf flow propagator} that embodies the effect of the polymer
chain upon the solution, is given by
\begin{eqnarray}
\Tind{}(\rvec,\rrvec) & = & - \int\!\!\!\int_0^L\! ds\, ds'
\, \delta (\rvec - \Rind{s}) \, {\bf T}_{ss'}\,
\delta (\rrvec - {\bf R}_{s'}) \nonumber\\
{\bf T}_{ss'} & = & \Tind{}(\Rind{s},\Rind{s'}) = \Ginv{ss'} -
\int\!\!\!\int_0^{L}\! dt\,dt' \, \Ginv{st} \cdot {\bf
g}^{-1}_T \cdot \Ginv{t's'}
\label{t_oper}\\
& & + \int\!\!\!\int_0^{L}\! dt\,dt' \,
\Ginv{st} \times \Sind{t} \cdot {\bf g}^{-1}_R
\cdot \Sind{t'} \times \Ginv{t's'} , \nonumber 
\end{eqnarray}
where, as previously mentioned, the subscripts $s,s',\dots$ represent the
arguments $\Rind{s},\Rind{s'},\dots$ of the operators. {\bf T} depends in
general upon the position, the structure and the geometry of the copolymer.

Finally, we obtain the equation for the macroscopic velocity field by averaging
Eq.~(\ref{sol_scatt}) with respect to the random distribution of $\Rind{0}$:
\begin{equation}
{\bf u}(\rvec) = \la{\bf v}(\rvec)\ra_0 = {\bf v}_0(\rvec) + {\bf G}\ast\la
{\bf T}\ra_0 \ast {\bf v}_0 |_{\rvec},
\label{sol_average}
\end{equation}
where we introduced the notations $\bbox{\ast}$ and $|_{\rvec}$ to denote the
convolution operation and its final argument: ${\bf A}\ast {\bf B}|_{\rvec} =
\int\!d\rrvec \, {\bf A}(\rvec,\rrvec)\cdot{\bf B}(\rrvec)$.

The connection with the experimentally measurable quantities is made by
introducing the self-energy tensor ${\bf \Sigma}(\rvec,\rrvec)$ defined by 
\begin{equation}
\int\! d\rrvec\, {\bf\Sigma}(\rvec,\rrvec)\cdot\la{\bf v}(\rrvec)\ra_0 =
{\bf\Sigma}\ast\la{\bf v}\ra_0 = \la\int_0^{L}\! ds \, \delta(\rvec -
\Rind{s})\sigind{s}\ra_0 .
\end{equation}
Next we average directly Eq.~(\ref{velocity_sol}) over $\Rind{0}$ and we
insert the expression of ${\bf\Sigma}$ from above, which yields
\begin{equation}
{\bf u}(\rvec) = {\bf v}_0(\rvec) + \int\!\!\!\int\! d\rrvec\,d\rrrvec \, {\bf
G}(\rvec - \rrvec)\cdot{\bf\Sigma}(\rrvec,\rrrvec)\cdot{\bf u}(\rrrvec) .
\label{sol_sigma}
\end{equation}
Taking into account that ${\bf v}_0(\rvec) = {\bf G}\ast{\bf F}|_{\rvec}$, we
can formally manipulate this equation to obtain
\begin{equation}  
{\bf u}(\rvec) = \bbox{\cal G}\ast{\bf F}|_{\rvec} = \int\! d\rrvec
\,\bbox{\cal G}(\rvec - \rrvec)\cdot {\bf F}(\rrvec) ,
\label{sol_effective}
\end{equation}
in which $\bbox{\cal G}$---the {\sf effective Oseen tensor}---is given by
\begin{eqnarray}
\calGind{}(\rvec,\rrvec) & = & \left(\openone \,-\, {\bf G} \ast
\bbox{\Sigma}\right)^{-1} \ast \left. {\bf
G}\rbarsub{(\rvec,\rrvec)}\nonumber\\
\calGind{}(\kvec)  & = & {\unit{k}\over \eta_0 k^2 - \Sigma_{\perp}(k)}. 
\label{Oseen_eff}
\end{eqnarray}
Here, by taking the inner-product with the projectors $\unit{k}$ and $\kk$ that
obey $(\unit{k})\cdot(\unit{k}) = \unit{k}$ and $\kk \cdot(\unit{k}) = 0$, we
decomposed the tensors in their transverse and longitudinal parts as in $ {{\bf
A} =} A_{\perp}(\unit{k}) + A_{\|} \kk$, with ${\bf A}$ a symmetric
tensor. After Fourier transforming and projecting the transversal and
longitudinal components of the tensors, we can calculate the self-energy as a
function of the flow propagator $\Tind{}$ by equating the expressions of ${\bf
u}(\rvec)$ from Eq.~(\ref{sol_average}) and Eq.~(\ref{sol_sigma}). These
operations yield
\begin{equation}
\bbox{\Sigma}(\kvec) = { \text{T}_{\perp}(k) \over 1 +
\text{G}_{\perp}(k)\text{T}_{\perp}(k)} \, (\unitkk) + \text{T}_{\|}(k) \, \kk
,  \label{sig_low}
\end{equation}
where $\Tind{}(\kvec)$ is the Fourier transform of $\la\Tind{}(\rvec -
\rrvec)\ra_0$ and $\text{G}_{\|} = 0$ (Eq.~(\ref{Oseen})). Note that the
equation for $\bbox{\Sigma}$ is exact. In the dilute limit we work with, we
can further approximate
\begin{equation}
{\bf \Sigma}(\kvec) \simeq \Tind{}(\kvec) ,
\label{sig_approx}
\end{equation}
where pre-averaging upon $\Tind{}(\kvec)$ is understood. 

From the structure of the effective Oseen tensor (\ref{Oseen_eff}) we identify
the change $\delta\eta_{\text{AB}}$ in the viscosity of the solution due to the
added copolymer, in the hydrodynamic limit $|\kvec| \to 0$, as
\begin{equation}
 {\eta_{\text{AB}}\over\eta_0} - 1 = - {1\over\eta_0} \, \lim_{k \to 0} \,
{\partial \over\partial k^2} \Sigma_{\perp}(k) .
\label{viscosity}
\end{equation}

\section{Calculation of $D_{\text{AB}}$, $\bz_{\text{AB}}$ and
$\delta\eta_{\text{AB}}$} 

In this section we present the detailed derivation of the transport
properties of the system. The essential aspect is the computation of the
structure factor, which involves, for the diblock copolymer case, the use of a
bivariate Gaussian probability distribution function. This calculation is
presented in Appendix~\ref{app:A}.  For any operator ${\bf A}(s,s')$ depending
on the arclength variables $s,s'$ we will use the following Fourier
representation (Rouse modes expansion in polymer language):
\begin{eqnarray}
{\bf A}(s,s') & = & {1\over L^2} \, \sum_{p=-\infty}^{\infty}\!\sum_{p' =
-\infty}^{\infty} {\bf A}_{p p'} \, e^{{2 i \pi\over L} p s - {2 i\pi\over L}
p' s'} \label{Fourier_ss} \\ 
{\bf A}_{p p'} & = & \int\!\!\!\int_0^L\! ds\,ds' \: {\bf A}(s,s') \, e^{- {2 i
\pi\over L} p s + {2 i\pi\over L} p' s'} ,\nonumber 
\eeqar
where we remind that $L$ is the total length of the polymer chain.  Also, the
inverse of ${\bf A}$ is defined by:
\begin{eqnarray}
\int_0^L\! ds' \, {\bf A}^{-1}(s,s')\cdot {\bf A}(s'.s'') & = & \delta(s - s'')
 \,\openone \label{Fourier_inverse}\\ 
\sum_{p'= -\infty}^{\infty} {\bf A}^{-1}_{p p'} \cdot {\bf A}_{p' p''} & = &
 L^2 \, \delta_{p p''}\, \openone \nonumber 
\eeqar

\subsection{Diffusion coefficient $D_{\text{AB}}$}

From Eqs.~(\ref{total_force}),(\ref{friction_coef}) we obtain
immediately---after preaveraging (which is equivalent to replacing all
instances of the Oseen tensor ${\bf G}(s,s')$ with its configurational average
$\la{\bf G}(s,s')\ra$)---that ${\bf D}_{\text{AB}}$, a tensor quantity in
general, is just the inverse of the friction coefficient $\bz_{\text{AB}}$
(Einstein relation). Thus we write
\begin{eqnarray}
{{\bf D}_{\text{AB}}\over k_B T} & = & D_{\text{AB}}\,\openone  = {1\over L^2}
\int\!\!\!\int_{0}^{L}\!ds\,ds'\,\la{\bf G}(s,s')\ra \label{diffusion_AB}\\ 
& = & {1\over L^2} \Gind{00} \;=\; {1\over 3 \pi^2\eta_0 L^2} \,
\int_{0}^{\infty} \!dk \int\!\!\!\int_{0}^{L}\!ds\,ds'\,\la \exp[i
\kvec\cdot({\bf S}_s - {\bf S}_s')]\ra \, \openone , \nonumber
\end{eqnarray}
where we used the Fourier expansions (\ref{Fourier_hydro}), (\ref{Fourier_ss})
and the integral $\int\!  d\Omega_k \, (\unit{k}) = {8 \pi\over 3}
\openone$. Splitting the double integral over the arclength variables $s$ and
$s'$ in the regions $[0,L_A];[L_A ,L]$, inserting the value of the exponential
term (the structure factor of the diblock copolymer) from App.~\ref{app:A} and
integrating over $k$, we get
\begin{equation}
{D_{\text{AB}}\over k_B T} = {16\over \pi \sqrt{\pi}}{1\over L^2
	}{1\over\eta_0} {1\over l_A l_B} \left[ \text{R}_A^3 \left({l_B\over
	l_A} - 1 \right) + \text{R}_B^3 \left({l_A\over l_B} - 1 \right) +
	\left(\text{R}_A^2 + \text{R}_B^2 \right)^{3/2} \right] ,
\label{diffusion}
\end{equation}
where $\text{R}_A = \sqrt{L_A l_A\over 6}$, $\text{R}_B = \sqrt{L_B l_B\over
6}$ are the radii of gyration of the two segments A and B, respectively, of the
copolymer. It is readily checked that $D_{\text{AB}}$ reduces to the classical
Kirkwood-Riseman (K-R) result \cite{KR}
\begin{equation}
	{D\over k_B T} = {1\over\zeta}  = {8\sqrt{2}\over 3} \text{h} = {4\over
	9\pi\sqrt{\pi}\eta_0 \text{R}_g},
\label{Kirkwood}
\end{equation}
with ${1\over\text{h}} = \eta_0 l \sqrt{12 \pi^3 N} = 6\pi\sqrt{2 \pi}\eta_0
\text{R}_g$ and $\text{R}_g = \sqrt{N l^2\over 6}$, in the limits $l_A = l_B$,
$N_A = 0$ or $N_B = 0$.  

The dimensionless variables that control the diffusion of the copolymer are
made apparent by normalizing the expression (\ref{diffusion}) for
$D_{\text{AB}}$ to the classical value (\ref{Kirkwood}) of $D$ of a polymer
with the length $L_N = N l_A$, where $N = N_A + N_B$. Defining $t = {l_B\over
l_A}$ and $x = {N_A\over N}$ yields, upon normalization, 
\begin{equation}
{D_{\text{AB}}\over D} = {t^2 (1 - x)^{3/2}(1 - t) - x^{3/2} (1 - t) + [x +
 t^2 (1 - x)]^{3/2}\over t [x + t (1 - x)]^2} ,
\label{diffusion_norm}
\end{equation}
which is the expression plotted in Fig.~\ref{fig:diffAB} as a function of $x$
and $t$, where $t$ has values both smaller and larger than $1$. When $t = 1$,
$x = 1$ or $x = 0$ we obtain the meaningful limits $1$, $1$ and ${1\over t}$,
respectively, the last result (when $x = 0$) representing just the diffusion
coefficient of a polymer of length $L_B = N l_B$ normalized to the
corresponding value of a chain with length $L_N = N l_A $. When $t = 0$, the
limit of (\ref{diffusion_norm}) is ${1\over\sqrt{x}}$. From
Fig.~\ref{fig:diffAB} we notice that a slight increase in the monomer fraction
of the A block in the A-B copolymer when $t = {l_B\over l_A} < 1$ has a much
stronger effect on $D_{\text{AB}}$ than an even large increase of $x$ in the
regime with $t > 1$.

A more relevant relation---particularly from an experimental point of
view---is obtained by computing the ratio between $D_{\text{AB}}$ and
the diffusion coefficient $D_A$ of the block A given by the standard result
(\ref{Kirkwood}) (with $N$ replaced by $N_A$), which immediately gives 
\begin{equation}
{D_{\text{AB}}\over D_A} = \sqrt{x}\, {t^2 (1 - x)^{3/2}(1 - t) - x^{3/2} (1 -
 t) + [x + t^2 (1 - x)]^{3/2}\over t [x + t (1 - x)]^2}.
\label{diffusion_normNA}
\end{equation}
This expression is plotted as a function of $x$ and $t$ in
Fig.~\ref{fig:diffABA} for a sequence of values of $t$ in the range
$[1/10-10]$, the points with $t>1$ being marked by circles joined by lines. Now
at each fixed value of $x$ (correspondingly $N_A$) one can compare the change
in the diffusion coefficient of the diblock copolymer with respect to the
diffusion constant of one of its components, which has direct implications in 
separation techniques.

We remark that at small $x$ values and $t\leq 1/4$ a maximum in
${D_{\text{AB}}\over D_A}$ occurs. Also, $D_{\text{AB}}$ exhibits a change in
curvature when the $t$ parameter goes from $t<1$ to $t>1$, which motivated us
in calculating the separation curves $\Delta (D_{\text{AB}}/D_A )$, defined as
the difference between two values of the ratio $D_{\text{AB}}/D_A$
corresponding to distinct values of $t$, as a function of $x$. These curves are
displayed in Fig.~\ref{fig:diffSEP}. Physically, the behavior of the diffusion
coefficient of the diblock copolymer opens the way to control the process of
separation of macromolecules by {\sf tagging} the targets with other
weight-controlled polymers. In particular, one can think of modifying the
diffusion coefficient of a given polymer chain A by attaching to it another
polymer B, with $t$ having a prescribed value. This is illustrated in
Fig.~\ref{fig:diffSEP} by the maxima occurring in the $\Delta
(D_{\text{AB}}/D_A )$ function. There is also a range of the $x$ values that
offers an increased {\sf efficiency} of the separation. This domain is much
more localized for $t <1$ than for $t>1$, but the absolute value of the peak of
$\Delta (D_{\text{AB}}/D_A )$ is highest when one compares---at the same $x$
ratio---two copolymers with $t$ greater and smaller than one. This is
reasonable---$t$ small means $l_A$ big, so even small variations in $x$ have
more significant effects than for pairs of copolymers with $t$ comparable to
one.

\subsection{Friction coefficient $\bz_{\text{AB}}$}

From Eq.~(\ref{friction_coef}), expanding the inverse Oseen tensor ${\bf
G}_{ss'}^{-1}$ in double Fourier series (\ref{Fourier_ss}), we find that
$\bz_{\text{AB}} = \Ginv{00}$.  Because the $\Gind{pp'}$ tensor, due to the
diblock structure of the copolymer with different Kuhn lengths, is not diagonal
in the $p$ index, the calculation of the friction coefficient is more
difficult. In general, one needs to decompose the $\Ginv{pp'}$ matrix in its
diagonal and non-diagonal parts and then to use a Born expansion up to the
desired order \cite{MF79}. Fortunately, next we will show that the $\Ginv{00}$
factor can be computed by direct numerical inversion.

We start with the expression of $\Gind{pp'}$ elements given by
Eq.~(\ref{Fourier_ss}):
\begin{equation}
\Gind{pp'} = \int\!{d\kvec \over (2\pi)^3} \, {\unit{k}\over\eta_0 k^2}
	\int\!\!\!\int_0^L\! ds\,ds'\, \la\exp[i \kvec\cdot(\Sind{s} -
	\Sind{s'})]\ra \exp[- {2 i \pi\over L} p s +{2 i \pi \over L} p' s' ]
\end{equation}
To evaluate the integrals, first we split the domain of integration
$(0,L)\times(0,L)$ over $s$ and $s'$ in four regions: $(0,L_A)\times (0,L_A)$,
$(0,L_A)\times(L_A,L)$, $(L_A,L)\times(0,L_A)$ and $(L_A,L)\times(L_A,L)$, and
then we insert the expression of the structure factor from
App.~\ref{app:A}. As the remaining integrations are rather intricate, we
present the computation of $\Gind{pp'}$ elements in
Appendix~\ref{app:B}. The final result is:
\begin{mathletters}
\label{Rouse_matrix}
\begin{eqnarray}
\Gind{pp}|_{p\neq 0} & = & L^2 \,{1\over \pi \sqrt{6\pi}\eta_0}\,{1\over\sqrt{L
|p|}} \, \left[ {L_A \over L} {1\over\sqrt{l_A}} + \left(1 - {L_A\over L}
\right){1\over \sqrt{l_B}} \right]\, \openone\\ 
\Gind{p p'}|_{p \neq 0} & = & L^2 \, {1\over 2
\pi^2\sqrt{6\pi}\eta_0}\,{1\over\sqrt{L |p|}}\, {1\over i (p' - p)} \left[
{1\over \sqrt{l_B}} - {1\over\sqrt{l_A}}\right]\, \left[1 - e^{{2\pi i L_A
\over L}(p' - p)}\right] \, \openone \;\; (p,p'\in {\Bbb Z}).
\end{eqnarray}
\end{mathletters}
From Eqs.~(\ref{diffusion_AB};\ref{diffusion}) the $\Gind{00}$ element is just
$L^2 {\bf D}_{\text{AB}}/(k_B T)$. The $\Gind{p p'}$ matrix is hermitian
($\Gind{p p'} = \Gind{p' p}^{\ast}$) and diagonally dominant, thus securing a
well behaved inverse. We remark that only when $l_A = l_B$, $L_A = 0$ or $L_A =
L$ the non-diagonal elements of $\Gind{pp'}$ vanish.

To further proceed with the numerical inversion, we need to reduce the
$\Gind{pp'}$ tensor to a dimensionless form. This task is accomplished by
noticing that the Rouse coefficients $\Gind{pp'}$ can be written as
\begin{equation}
\Gind{pp'} = L^2 \, {1\over 4\pi^2\sqrt{6 \pi}\eta_0}\, {1\over\sqrt{L l_A}} \,
{\bf K}_{pp'}(x,t) ,
\label{Oseen_K}
\end{equation}
where ${\bf K}$ is a purely numerical tensor, function of the dimensionless
parameters $x = {N_A\over N}$ and $t = {l_B\over l_A}$ and with the matrix
representation (hermitian) given by
\begin{mathletters}
\begin{eqnarray}
{\bf K}_{0 0}(x,t) & = & {32 \pi\over 3} \, {t^2 (1 - x)^{3/2}(1 - t) - x^{3/2}
(1 - t) + [x + t^2 (1 - x)]^{3/2}\over t [x + t (1 - x)]^{3/2}}\,
\openone\\ 
{\bf K}_{p p}(x,t)|_{p\neq 0} & = & {4\pi\over \sqrt{|p|}}\, \left[y + {1 -
y\over \sqrt{t}} \right]\,\openone ;\;\;\; y = {x \over x + t (1 - x)}\\ 
{\bf K}_{p p'}(x,t)|_{p\neq 0} & = & {2\over i (p' - p)} \, {1\over
\sqrt{|p|}}({1\over \sqrt{t}} - 1)[1 - e^{2 i \pi (p'-p) y}] \, \openone ,
\end{eqnarray}
\end{mathletters}
where we employed $L = L_A + L_B = N l_A (x + t (1 - x))$. Note that for our
choice of $t$ the limit of $t\to 0$ for ${\bf K}_{pp'}$ is not immediate. Next,
applying the definition (\ref{Fourier_inverse}) we construct the inverse
$\Ginv{pp'}$ as
\begin{equation}
\sum_{p'=-\infty}^{+\infty} \Ginv{pp'}\cdot\Gind{p'p''} =
 L^2 \delta_{pp''}\,\openone 
\end{equation}
and substituting $\Gind{pp'}$ from Eq.~(\ref{Oseen_K}), we find
that
\begin{equation}
\Ginv{pp'} = 4 \pi^2 \sqrt{6 \pi} \eta_0 \sqrt{L l_A} \, \Kinv{pp'}(x,t) ,
\label{Oseen_inv_K}
\end{equation}
where the $\Kinv{}$ operator is defined by the equation:
\begin{equation} 
\sum_{p'= -\infty}^{\infty} {\bf K}^{-1}_{p p'}(x,t) \cdot {\bf K}_{p'
p''}(x,t) = \delta_{p p''}\, \openone .
\label{K_Inv_Def} 
\end{equation}
\indent Eventually, recalling that $\bz_{\text{AB}} = \Ginv{00}$ and after
normalizing to the friction coefficient $\zeta = 9\pi\sqrt{\pi}\eta_0
\text{R}_g/4$ (see Eq.~(\ref{Kirkwood})) of a homogeneous polymer of length $L
= N l_A$, we obtain
\begin{equation}
{\zeta_{\text{AB}}\over\zeta}(x,t) = {32 \pi\over 3} \, \sqrt{x + t (1 - x)} \,
\text{K}_{00}^{-1}(x,t) ,
\label{friction_result}
\end{equation}
where we dropped the unit dyad from the formulas, as inessential. The limiting
cases when the tensor $\Kinv{pp'}$ is diagonal can be readily checked
analytically. We then get:
\begin{eqnarray}
\text{K}_{00}(x,1) & = & {32 \pi\over 3}; \;\;\;\;\;\; \text{K}_{pp}(x,1) =
{4\pi\over \sqrt{|p|}}\nonumber\\ 
\text{K}_{00}(0,t) & = & {32 \pi\over 3}{1\over\sqrt{t}};\;\;
\text{K}_{pp'}(x,1) = 0 \;\;\text{for}\;\; p\neq p' 
\label{K_elements}\label{K_limit}\\ 
\text{K}_{00}(1,t) & = & {32 \pi\over 3}; \;\;\;\;\;\; \text{K}_{pp}(x,0) = {32
\pi\over 3}.\nonumber
\end{eqnarray}
For these limits the inverse of $\text{K}_{00}$ is just $\text{K}_{00}^{-1} =
(\text{K}_{00})^{-1}$ and the friction coefficient
${\zeta_{\text{AB}}\over\zeta}(x,t)$ reduces accordingly to the physically
correct values:
\begin{equation} 
{\zeta_{\text{AB}}\over\zeta}(x,1) = 1; \;\;\;
{\zeta_{\text{AB}}\over\zeta}(0,t) = t; \;\;\;
{\zeta_{\text{AB}}\over \zeta}(1,t) = 1 .
\end{equation}

The explicit inversion of $\text{K}_{00}(x,t)$ is carried out by first fixing
the values of the parameters $x$ and $t$. Then we iterate the dimensions of the
$\text{K}_{p p'}(x,t)$ matrix and we numerically invert it for each $|p| =
\overline{1,p_{\text{max}}}$ until $\text{K}_{00}^{-1}(x,t)$ converges to the
desired precision---we have chosen $10^{-7}$ for computational convenience---at
a certain $p_{\text{max}}$.

The final results for ${\zeta_{\text{AB}}\over\zeta}$ calculated from
Eq.~(\ref{friction_result}) have been plotted in Figs.~\ref{fig:frLess} and
\ref{fig:frBig} against the fraction $x$ of $A$ beads, for values of $t$
smaller and greater than $1$. Points where a numerical evaluation was carried
out are marked by symbols, their shape corresponding  to different
values of $t$.

 An illuminating aspect of the result is that neglecting the off-diagonal
components of the Rouse tensor $\Gind{pp'}$ and inverting it directly (then
$\bz_{\text{AB}}$ is just $L^2 (\Gind{00})^{-1}$) still leads to the
qualitatively correct behavior of $\bz_{\text{AB}}$, as displayed by the
continuous lines in Figs.~[\ref{fig:frLess}, \ref{fig:frBig}].

\subsection{Intrinsic Viscosity $[\eta_{\text{AB}}]$}

As seen from Eq.~(\ref{viscosity}), the information regarding the change in the
viscosity of the solution is contained in the self-energy
$\bbox{\Sigma}(\kvec)$. To find it we proceed by averaging over the
random position ${\bf R}_0$ of the center-of-mass of the polymer and Fourier
transforming Eq.~(\ref{t_oper}), which yields
\begin{eqnarray}
{\bf\Sigma}(\kvec) \simeq \Tind{}(\kvec) & = & -{1\over V}\, \int\!d(\rvec -
\rrvec)\int\!d{\bf R}_0\, \exp[i \kvec\cdot(\rvec - \rrvec)] \nonumber\\ 
& & \times \int\!\!\!\int_0^L\!  ds\,ds'\, \delta(\rvec - {\bf R}_0 - \Sind{s})
\Tind{ss'} \delta(\rrvec - {\bf R}_0 - \Sind{s'})\\ 
& & = - {1\over V} \int\!\!\!\int_0^L\! ds\,ds'\, \Tind{ss'} \exp[i
\kvec\cdot(\Sind{s} - \Sind{s'})], \nonumber
\end{eqnarray}
where $V$ is the volume of the suspension.  Inserting the full form of
$\Tind{ss'}$ from (\ref{t_oper}) produces
\begin{equation}
{\bf\Sigma}(\kvec) = - {1\over V} [\Tind{}^{(1)}(\kvec) - \Tind{}^{(2)}(\kvec)
+ \Tind{}^{(3)}(\kvec)] ,
\label{sig_viscos}
\end{equation}
with the new factors being given by:
\begin{mathletters}
\label{T_comp}
\begin{eqnarray}
\Tind{}^{(1)}(\kvec) & = & \int\!\!\!\int_0^L\! ds\,ds' \, \Ginv{ss'} \,
\exp[i\kvec\cdot(\Sind{s} - \Sind{s'})] \label{T1}\\ 
\Tind{}^{(2)}(\kvec) & = & \int\!\!\!\int\!\!\!\int\!\!\!\int_0^L\!
ds\,ds'\,dt\,dt'\, \Ginv{st}\cdot {\bf g}_T^{-1} \cdot \Ginv{t's'} \,
\exp[i\kvec\cdot(\Sind{s} - \Sind{s'})] \label{T2}\\ 
\Tind{}^{(3)}(\kvec) & = & \int\!\!\!\int\!\!\!\int\!\!\!\int_0^L\!
ds\,ds'\,dt\,dt'\, \Ginv{st}\times \Sind{t} \,\cdot\, {\bf g}_R^{-1} \,\cdot\,
\Sind{t'}\times \Ginv{t's'} \, \exp[i\kvec\cdot(\Sind{s} - \Sind{s'})]
\label{T3} 
\end{eqnarray}
\end{mathletters}
Here, we recall that $\Ginv{ss'}$ is defined by Eq.~(\ref{Oseen_inverse}) and
should actually read  $\Ginv{}(\Sind{s} - \Sind{s'})$. From (\ref{coeffs}) we
calculate---within the pre-averaging approximation---the frictional
coefficients ${\bf g}_T$ and ${\bf g}_R$ as
\begin{mathletters}
\label{rotfric_coeff}
\begin{eqnarray}
\la{\bf g }_T\ra & = & \int\!\!\!\int_0^L\! ds\,ds'\, \la\Ginv{ss'}\ra =
\Ginv{00} \label{fric_viscos}\\ 
\la{\bf g}_R\ra & = & - \int\!\!\!\int_0^L\!  ds\,ds'\,\la\Ginv{ss'}\ra \cdot
\la \Sind{s}\times \openone \times \Sind{s'} \ra \label{rot_viscos}\\ 
& = & {2\over 3} \int\!\!\!\int_0^L\! ds \,ds'\, \la \Ginv{ss'}\ra\cdot \la
\Sind{s}\cdot\Sind{s'}\ra = {2\over 3} {1\over L^2} \sum_{p =
-\infty}^{+\infty} \sum_{p'=-\infty}^{+\infty} \Ginv{pp'}\cdot
\subs{F}{pp'}^{\ast} ,\nonumber
\end{eqnarray}
\end{mathletters}
where we used $\la\text{S}^{(i)}_s \, \text{S}^{(j)}_{s'} \ra = {1\over 3}
\la\Sind{s}\cdot\Sind{s'}\ra\delta_{ij}$ that implies
$\la(\Sind{s}\cdot\rvec)(\Sind{s'}\cdot\rvec)\ra = {1\over 3} r^2
\la\Sind{s}\cdot\Sind{s'}\ra$ for any vector $\rvec$. Note that the double
summation over $p$ and $p'$ is equivalent to taking the trace of
$\Ginv{}\cdot\subs{F}{}$. We also developed all quantities in the double
Fourier series (\ref{Fourier_ss}) and (\ref{Fourier_inverse}). The $\subs{F}{}$
elements are computed from
\begin{equation}
\subs{F}{pp'} = \text{F}_{pp'}\openone = \int\!\!\!\int_0^L\! \la\Sind{s}\cdot
\Sind{s'}\ra \, \exp[ -{2 i \pi \over L} p s + {2 i\pi\over L} p's'] \,
\openone .
\label{F_tensor}
\end{equation}
We remark that $\subs{F}{}$ is a hermitian operator.  Averaging
$\Sind{s}\cdot\Sind{s'}$ over the distribution of the segments $s$ and $s'$ for
the diblock copolymer is a rather involved operation and the details are
presented in Appendix~\ref{app:SS}.

Next we can simplify the calculations by taking directly the $k^2$ components
of $\Tind{}^{(i)}(\kvec)$ in (\ref{T_comp}). Then $\Tind{}^{(1)}$ becomes, upon
preaveraging, 
\cite{M81} 
\begin{eqnarray}
\Tind{k^2}^{(1)} & = & \int\!\!\!\int_0^L\! ds\,ds' \, \la\Ginv{ss'}\ra \cdot
\la (\kvec\cdot\Sind{s})(\kvec\cdot\Sind{s'})\ra \label{T1_final}\\
& = & {1\over 3} k^2 \, \int\!\!\!\int_0^L\! ds\,ds' \,
\la\Ginv{ss'}\ra \cdot \la\Sind{s}\cdot\Sind{s'}\ra  = {1\over 2} k^2 \,
\la{\bf g}_R\ra ,\nonumber
\end{eqnarray}
where ${\bf g}_R$ is given by Eq.~(\ref{rot_viscos}). 

In a similar manner, it follows that $\Tind{k^2}^{(2)}$ can be written as
\begin{eqnarray}
\Tind{k^2}^{(2)} & = & {1\over 3} k^2 \,
\int\!\!\!\int\!\!\!\int\!\!\!\int_0^L\!  ds\,ds'\,dt\,dt'\, \la\Ginv{st}\ra
\cdot \la{\bf g}_T^{-1}\ra  \cdot \la\Ginv{t's'}\ra
\la\Sind{s}\cdot\Sind{s'}\ra \label{T2_final}\\
& = & {1\over 3} {1\over L^2} \, k^2 \, \la{\bf g}_T^{-1}\ra  \sum_{pp'}
\Ginv{0p}\cdot \subs{F}{pp'}\cdot \Ginv{p'0} \, , \nonumber 
\end{eqnarray}
where the Fourier expansion (\ref{Fourier_ss}), (\ref{Fourier_inverse}) were
once again utilized. For a homogeneous polymer ($t = 1$), $\Ginv{pp'}$ becomes
diagonal and $\subs{F}{00} = 0$ (App.~\ref{app:SS}) so $\Tind{k^2}^{(2)}$
vanishes as required \cite{M81}.

Finally, the $k^2$ part of $\Tind{}^{(3)}(\kvec)$ results from 
\begin{eqnarray}
\Tind{k^2}^{(3)} & = & \la{\bf g}_R^{-1}\ra
\int\!\!\!\int\!\!\!\int\!\!\!\int_0^L\!  ds\,ds'\,dt\,dt'\, \la\Ginv{st}\ra
\cdot (\openone \times \la \kvec\cdot\Sind{s}\, \Sind{t}\ra) \nonumber\\
& & \cdot \,( \la\kvec\cdot\Sind{s'} \, \Sind{t'}\ra \times \openone) \cdot
\la\Ginv{t's'}\ra \\
& = & - {1\over 9} \la{\bf g}_R^{-1}\ra k^2 \, (\unit{k}) \cdot
\int\!\!\!\int\!\!\!\int\!\!\!\int_0^L\!  ds\,ds'\,dt\,dt'\,
\la\Ginv{st}\ra \nonumber\\
& & \cdot \la\Sind{t}\cdot\Sind{s}\ra \cdot \la
\Sind{s'}\cdot\Sind{t'}\ra \cdot \la\Ginv{t's'}\ra \nonumber
\end{eqnarray}
where we pulled out the ${\bf g}_R^{-1}$ factor as it is just a number
multiplying the unit tensor and we applied the relation $(\openone\times{\bf
a})\cdot({\bf a}\times\openone) = - a^2 (\unit{a})$, valid for any vector ${\bf
a}$.  Developing all factors in double Fourier (Rouse modes) series we obtain
\begin{equation}\
\Tind{k^2}^{(3)} = - {1\over 9} {1\over L^4} \la{\bf g}_R^{-1}\ra k^2 \,
(\unit{k}) \cdot \sum_{pq} \Ginv{pq}\cdot \subs{F}{pq}^{\ast} \cdot
\sum_{p'q'}\Ginv{q'p'}\cdot \subs{F}{q'p'}^{\ast} ,
\end{equation}
in which all the summation indices run over the $(-\infty,+\infty)$
interval. Replacing the sums in terms of the $\la{\bf g}_R\ra$ coefficient
(Eq.~(\ref{rot_viscos})) and noticing that in our approximation ${\la {\bf
g}_R^{-1}\ra \cdot\la {\bf g}_R\ra = 1}$, we get the simpler form
\begin{equation}
\Tind{k^2}^{(3)} = - {1\over 4} \la {\bf g}_R \ra \, k^2 \, (\unit{k}) \, .
\label{T3_final}
\end{equation}
\indent Collecting all the components of $\Tind{k^2}$ from
Eqs.~(\ref{T1_final}), (\ref{T2_final}) and (\ref{T3_final}) and inserting them
in Eq.~(\ref{sig_viscos}), we find that
\begin{eqnarray}
{\bf\Sigma}_{k^2}(\kvec) & = & - {1\over V} \left[{1 \over 2} k^2 \, \la
\text{g}_R\ra \openone - {1\over 3} k^2 \, \la\text{g}_T^{-1}\ra {1\over L^2}
\sum_{p=- \infty}^{+\infty}\sum_{p'=-\infty}^{+\infty}
\text{G}^{-1}_{0p}\text{F}_{pp'} \text{G}^{-1}_{p'0}\,\openone  \right. \\
& & - \left. {1\over 4} k^2 \, \la\text{g}_R\ra (\unit{k}) \right] \nonumber \,
, 
\end{eqnarray}
where we used the fact that all quantities are scalar multiples of the unit
dyad $\openone$ (e.g. $\la\text{g}_R\ra$ and $\la\text{g}_T^{-1}\ra$ are now
some functions to be computed and $\text{G}^{-1}_{pp'}$, $\text{F}_{pp'}$ are
the matrix elements of the Fourier representation of $\text{G}^{-1}$ and
$\text{F}$). The change in viscosity follows from Eq.~(\ref{viscosity}) by
applying the transverse projector $\unit{k}$ to the previous expression:
\begin{equation}
{\delta\eta_{\text{AB}}\over \eta_0}(\unit{k}) = {1\over V \eta_0} \left[ {1
	\over 4} k^2 \la\text{g}_R\ra - {1\over 3} k^2 \, \la\text{g}_T^{-1}\ra
	{1\over L^2} \sum_{p,p'} \text{G}^{-1}_{0p}\text{F}_{pp'}
	\text{G}^{-1}_{p'0} \right](\unit{k}) .
\label{viscosity_formal}
\end{equation}
Introducing the concentration of the copolymer $c_{\text{AB}} =
{M_{\text{AB}} \over V N_{\text{AV}}}$, with $N_{\text{AV}}$ Avogadro's
number and $M_{\text{AB}}$ the molecular mass, we calculate the intrinsic
viscosity $[\eta_{\text{AB}}]$ from
\begin{equation}
[\eta_{\text{AB}}] = {\eta_{\text{AB}} - \eta_0\over c_{\text{AB}}\eta_0} =
{\delta\eta_{\text{AB}}\over \eta_0} {V N_{\text{AV}}\over M_{\text{AB}}}.
\label{visc_intrinsic}
\end{equation}

Formally, we have the desired expression. Practically, we need to reduce it to
a dimensionless form amenable to numerical evaluation. To this end, we write
$\Ginv{}$ in terms of the $\Kinv{}$ operator from Eq.~(\ref{Oseen_inv_K})
obeying also (\ref{K_Inv_Def}), with $\subs{K}{}(x,t)$ introduced in
Eq.~(\ref{Oseen_K}). Then we express $\subs{F}{pp'}$ as
\begin{equation} 
\subs{F}{pp'} = {(N l_A)^4 \over 2 N}f_{pp'}(x,t) \,\openone\, ,
\label{F_dimless}
\end{equation}
where $x={N_A\over N}$, $t = {l_B\over l_A}$ and $f_{pp'}$ is a dimensionless
function of $x$ and $t$ (see App.~\ref{app:SS}). Inserting back in
Eq.~(\ref{rot_viscos}), we obtain the rotational friction coefficient as
\begin{equation}
\la\text{g}_R\ra = 4 \pi^2 \sqrt{2\pi\over 3}\eta_0 (N l_A^2)^{3/2} {1\over [x
+ (1 - x) t]^{3/2}} \sum_{p,p' = -\infty}^{+\infty} \text{K}_{pp'}^{-1}(x,t)
f_{pp'}^{\ast}(x,t)
\label{rot_dimless} 
\end{equation}
To check the result we quote some particular values of the $f_{pp'}(x,t)$
function:
\begin{eqnarray}
f_{00}(x,1) = 0 ; & f_{00}(0,t) = 0 &; f_{00}(1,t) = 0
\label{f_elements}\label{f_limits}  \\
f_{pp}(x,1) = {1 \over \pi^2 p^2} ; & f_{pp'}(x,1) = {1 \over 2 \pi^2 p p'} & ;
f_{pp}(x,0) = {x^3 \over \pi^2 p^2} .\nonumber
\end{eqnarray}
When $t = 1$ (the homogeneous case), the $\text{K}_{pp'}$ matrix is
diagonal, $\text{K}^{-1} = (\text{K})^{-1}$ and using (\ref{K_elements}) $\la
\text{g}_R\ra$ reduces to the known result
\begin{equation}
\la\text{g}_R\ra|_{t = 1} = {24\over\sqrt{\pi}} \eta_0 \left({N l_A^2\over
6}\right)^{3/2} \sum_{p=1}^{\infty} {1\over p^{3/2}}.
\end{equation}
Now, as $\la\text{g}_T^{-1}\ra$ is just the inverse of the translational
friction coefficient $\la\text{g}_T\ra$ given by (\ref{fric_viscos}), we can
calculate
\begin{eqnarray}
\la\text{g}_T\ra & = & D_{\text{AB}} = {8\over 3\pi\sqrt{6 \pi}\eta_0}
{1\over\sqrt{N l_A^2}} \label{fric_dimless}\\
& \times & {t^2 (1 - x)^{3/2}(1 - t) - x^{3/2} (1 - t) + [x +
 t^2 (1 - x)]^{3/2}\over t [x + t (1 - x)]^2} \nonumber ,
\end{eqnarray}
from which we recover the Kirkwood-Riseman result when $t\to 1$, as already
discussed.

Going back to the viscosity formula (\ref{viscosity_formal}), we use
Eqs.~(\ref{Oseen_inv_K}), (\ref{F_dimless}), (\ref{rot_dimless}) and
(\ref{fric_dimless}) to express everything in dimensionless variables and,
after normalizing to the non-free draining result \cite{M81,KR}
\begin{equation} 
{\delta\eta\over\eta_0} = {1\over V}{6\over\sqrt{\pi}}\left({N l_A^2\over
6}\right)^{3/2} \text{Zeta}(3/2) ,
\end{equation}
where $\text{Zeta}(3/2) = \sum_{p = 1}^{\infty} {1\over p^{3/2}}$ (the Riemann
function), we finally obtain
\begin{eqnarray}
{\delta\eta_{\text{AB}}\over \delta\eta} (x,t) & = & {2\pi^3\over
	\text{Zeta}(3/2)} {1\over [x + t (1 - x) ]^{3/2}}\left[
	\sum_{p,p'=-\infty}^{+\infty} \text{K}_{pp'}^{-1
	}(x,t)f_{pp'}^{\ast}(x,t) \right. \nonumber\\ 
	& & - \left. {64\pi\over 3} {t^2 (1 - x)^{3/2}(1 - t) - x^{3/2} (1 - t)
	+ [x + t^2 (1 - x)]^{3/2}\over t [x + t (1 - x)]^{3/2}}
	\right. \label{visc_dimless}\\  
	& & \times \left. \sum_{p,p'=-\infty}^{+\infty} \text{K}^{-1}_{0p}(x,t)
	f_{pp'}(x,t) \text{K}^{-1}_{p'0}(x,t) \right]. \nonumber
\end{eqnarray}
Once again, we can check analytically some limits of this expression
by using the matrix elements $\text{K}_{pp'}$ and $f_{pp'}$ from
Eqs.~(\ref{K_elements}) and (\ref{f_elements}):
\begin{equation}
\begin{array}{lcr}
{\delta\eta_{\text{AB}}\over \delta\eta} (x,1) = 1; &
{\delta\eta_{\text{AB}}\over \delta\eta} (1,t) = 1; &
{\delta\eta_{\text{AB}}\over \delta\eta} (0,t) = t^3 .
\end{array}
\end{equation}
To compare directly the increase in viscosity due to the A-B copolymer to the
change due to the polymer A only (computed from the K-R result), we only need
to multiply (\ref{visc_dimless}) with ${1\over x^{3/2}}$, which then yields
\begin{equation}
{\delta\eta_{\text{AB}}\over \delta\eta_A}(x,t) = {1\over x^{3/2}}
{\delta\eta_{\text{AB}}\over \delta\eta} (x,t) 
\label{viscABA_dimless}
\end{equation}

The relative intrinsic viscosity of the suspension follows readily from
(\ref{visc_intrinsic}):
\begin{equation}
{[\eta_{\text{AB}}]\over [\eta]} = {M_{\text{AB}}\over M}
{\delta\eta_{\text{AB}}\over \delta\eta} (x,t),
\end{equation}
where $M$ and $[\eta]$ are the molecular mass and the intrinsic viscosity of a
solution of a polymer with $N=N_A+N_B$ beads and $l_A$ Kuhn-length.

In Fig.~\ref{fig:viscosity} the viscosity calculated from (\ref{visc_dimless})
is plotted against $x = {N_A\over N}$ for different values of the parameter $t
= {l_B\over l_A}$.  From a practical point of view, it is reassuring that,
checking once again the validity of the Kirkwood-Riseman approximation by
computing $\delta\eta_{\text{AB}}$ neglecting all the non-diagonal elements of
$\Gind{pp'}$, we obtain a very good agreement with the exact numerical
evaluations. This aspect is illustrated in Fig.~\ref{fig:viscosity}, where the
continuous lines represent the calculation done using only the diagonal
elements $\Gind{pp}$ only.

Based on this finding, in Fig.~\ref{fig:viscosityABA} the viscosity calculated
from (\ref{visc_dimless}) but employing the Kirkwood-Riseman approximation is
plotted against $x$ at various $t$ values.  Similar to the observations in the
diffusion case, we notice that the intrinsic viscosity
$[\eta_{\text{AB}}]/[\eta_A]$ attains a minimum matching the maximum of
$D_{\text{AB}}/D_A$ from Fig.~\ref{fig:diffABA}. This should be useful for some
technological processes where a control of the viscosity of a solution is
desired. 

\section{Conclusions}

In this paper we have considered the stationary dynamics of an infinitely
diluted solution of freely translating and rotating Gaussian diblock
copolymers. The copolymer chain consists of two components A and B, with
Kuhn-lengths $l_A$ and $l_B$ and number of segments $N_A$ and $N_B$,
respectively. There is no excluded-volume interaction and the solution is
described by the linearized Navier-Stokes equation. The hydrodynamics is
coupled to the chain dynamics by the no-slip boundary condition.

Extending to our case the cluster expansion theory\cite{FM78,MF82,M82,FM82,RPM}
previously used to study the hydrodynamics of suspensions of polymers and
spheres, we developed explicitly the necessary theoretical elements and we
obtained analytically and calculated numerically the following physical
quantities of interest: the diffusion coefficient $D_{\text{AB}}$, the friction
coefficient $\bz_{\text{AB}}$ and the intrinsic viscosity
$[\eta_{\text{AB}}]$. The results have been displayed graphically as functions
of the dimensionless variables characterizing the problem, $x = {N_A\over N}$
and $t = {l_B\over l_A}$.

Even for the simple system studied here, an interesting behavior of
$D_{\text{AB}}$ and implicitly of the other quantities, emerged. The main
finding of the present work is that by adjusting $x$ (the molecular mass
ratio) and $t$ parameters we can control the process of separation of diblock
copolymers (or, more generally, combined {\sf objects}) under hydrodynamic
flow.  To illustrate this aspect we constructed separation curves by taking the
difference $\Delta(D_{\text{AB}}) = D_{\text{\/AB}}(x,t_1) -
D_{\text{\/AB}}(x,t_2)$ and representing it as a function of $x$ and the pair
$t_1-t_2$, which allows us to identify both the absolute maxima and the ranges
of $x$ that maximize the difference $\Delta(D_{\text{AB}})$ for given $t_1$ and
$t_2$. As an immediate extension of the present work, an application of our
approach to the process of electrophoretic separation is in progress.

The formalism presented is applicable in more general contexts not limited to
Gaussian chains. For example, when excluded-volume or Coulombic interactions
are present we can capture their effects---in certain conditions---by
renormalizing the Kuhn-lengths $l_A$ and $l_B$ to some {\sf effective}
values. As an additional observation, the usual Kirkwood-Riseman approximation
is not necessary. But, rather surprisingly, our calculations have shown
rigorously that one can still safely apply it.

\acknowledgments

Acknowledgment is made to the Materials Research Science and Engineering Center
at the University of Massachusetts, and the NSF Grant DMR 9625485.

\appendix

\section{Computation of $\la\exp[\lowercase{i}
\kvec(\Sind{\lowercase{s}}-\Sind{\lowercase{s'}})] \ra$} 
\label{app:A}

Consider a Gaussian polymer chain consisting of $N$ links each of mean-square
length $l_k$, with $k\in\overline{1,N}$. Following Yamakawa,\cite{YAM} we can
construct the probability distribution function $P(\Sind{i} - \Sind{j})$ of the
distance between two beads $i$ and $j$ located at $\Sind{i}$ and $\Sind{j}$,
respectively, about the center-of-mass of the chain as
\begin{equation}
P(\Sind{i} - \Sind{j}) = \left({3\over 2 \pi \la l^2 \ra}\right)^{3/2}
\exp\left[- {3  |\Sind{i}-\Sind{j}|^2  \over 2 \la l^2\ra}\right] ,
\end{equation}
with $\la l^2\ra = \sum_{m\geq j+1}^{i} l_k^2$. Then, $\la\exp[i
\kvec\cdot(\Sind{i} - \Sind{j})]\ra$ is the generating function of the moments
of the $P(\Sind{i}-\Sind{j})$ distribution and is given by
\begin{equation}
M[\kvec] = \la\exp[i
\kvec\cdot(\Sind{i} - \Sind{j})]\ra = \exp(- {1\over 6}k^2 \la l^2 \ra).
\end{equation}
In our problem we have only two types of links, A and B, with average lengths
$l_A$ and $l_B$ and the calculation of the structure factor averaged over the
distribution of beads yields:
\begin{equation}
\la e^{i \kvec\cdot({\bf S}_s - {\bf S}_{s'})} \ra \; = \; \left\{
\begin{array}{ll}
e^{-{1\over 6}\,k^2 |s - s'| l_{A,B}} & \text{if $s,s' < L_A$ or $s,s' >
L_A$};\\  
e^{- {1\over 6}\,k^2 [(L_A - s)l_A + (s' - L_A) l_B]} & \text{if $s < L_A$ and
$s' > L_A$}; \\
e^{- {1\over 6}\, k^2 [(L_A - s')l_A + (s - L_A)l_B]} & \text{if $s > L_A$ and
$s' < L_A$};
\end{array} \right. 
\label{structure_elem}
\end{equation}

In the expressions above we introduced the arclength variables $s$ and $s'$
instead of the numbers $i$, $j$ (e.g. $s = {i l_A}$ if $s < L_A$) and $L_A = 
N_A l_A ;\; L_B = N_B l_B$.
 
\section{Computation of the matrix elements
$\Gind{\lowercase{p}\lowercase{p'}}$} 
\label{app:B}

As previously shown, the Fourier coefficients of the Oseen tensor are given by
\begin{eqnarray}
\Gind{pp'} & = & \int\!{d\kvec \over (2\pi)^3} \, {\unit{k}\over\eta_0 k^2}
	\int\!\!\!\int_0^L\! ds\,ds'\, \label{app_B_Rouse}\\
& \times & \la\exp[i \kvec\cdot(\Sind{s} -
	\Sind{s'})]\ra \exp[- {2 i \pi\over L} p s +{2 i \pi \over L} p' s'
	]\nonumber 
\end{eqnarray}
Introducing the structure factor from (\ref{structure_elem}) and performing the
angular integral utilizing $\int\!d\Omega_k \, (\unit{k}) = {8\pi\over
3}\openone$,  we arrive at
\begin{eqnarray}
\Gind{pp'} & = & {1\over 3\pi^2 \eta_0} \int_0^{\infty}\! dk \, \left[
	\int\!\!\!\int_0^{L_A}\! ds \, ds' \, e^{-{1\over 6}\,k^2 |s - s'| l_A}
	\right. \nonumber\\
& + & \int\!\!\!\int_{L_A}^{L}\! ds \, ds' \, e^{-{1\over 6}\,k^2 |s - s'|
	l_B} \nonumber\\ 
& + & \int_0^{L_A}\!\!\!\int_{L_A}^{L}\! ds \,
	ds' \, e^{- {1\over 6}\,k^2 [(L_A - s)l_A + (s' - L_A) l_B]} \\
& + & \left. \int_{L_A}^{L}\!\!\int_{0}^{L_A}\! ds \, ds' \, e^{- {1\over 6}\,
	k^2 [(L_A - s')l_A + (s - L_A)l_B]} \right] \nonumber \\
& & \times \exp\left[- {2 i \pi\over L} p s +{2 i \pi \over L} p' s' \right] \,
	\openone .\nonumber
\end{eqnarray}
The evaluation of the integrals goes as follows. First we compute the
$\Gind{00}$ factor:
\begin{equation}
\Gind{00} = {16\over \pi \sqrt{\pi}\,\eta_0}
	{1\over l_A l_B} \left[ \text{R}_A^3 \left({l_B\over l_A} - 1 \right) +
      \text{R}_B^3 \left({l_A\over l_B} - 1 \right) + \left(\text{R}_A^2 +
	\text{R}_B^2 \right)^{3/2} \right]\openone \, ,
\end{equation}
where $\text{R}_{A,B} = (L_{A,B} l_{A,B}/6)^{1/2}$ are the radii of gyration of
the A and B components of the copolymer and the integral
$\int_0^{\infty}\!dx\,[{1\over x^2} - {1\over x^4} (1 - e^{-x^2})] =
{2\sqrt{\pi}\over 3}$ was used.

Next, making $p = p'\neq 0$ and integrating over $s$ and $s'$ yields
\begin{eqnarray}
\Gind{pp} & = & {1\over 3\pi^2 \eta_0} \int_0^{\infty}\! dk \,\left\{ {2 L_A^2
	\, (k \text{R}_A)^2 \over (k \text{R}_A)^4 + \left({2\pi L_A \over L}
	p\right)^2} + {2 (L - L_A)^2 (k \text{R}_B)^2 \over (k \text{R}_B)^4 +
	\left({ 2 \pi (L - L_A)\over L} p \right)^2} \right. \nonumber\\ 
& & + \left. {L_A (L - L_A) \over [(k \text{R}_A)^2 - {2\pi L_A \over L} p] [(k
	\text{R}_B)^2 - {2\pi (L - L_A) \over L} p]} [1 - \exp({2 i \pi
	L_A\over L} p - k^2 \text{R}_A^2)] \right.\nonumber\\ 
& & \times \left. [1 - \exp(- {2 i \pi L_A\over L} p - k^2 \text{R}_B^2)]
\right. \\
& & + \left. {L_A (L - L_A) \over [(k \text{R}_A)^2 + {2\pi L_A \over L} p] [(k
	\text{R}_B)^2 + {2\pi (L - L_A) \over L} p]}  [1 - \exp(-{2 i \pi
	L_A\over L} p - k^2 \text{R}_A^2)] \right. \nonumber\\
& & \times \left. [1 - \exp({2 i \pi L_A\over L} p - k^2 \text{R}_B^2)]
	\right\}\,\openone . \nonumber
\end{eqnarray}
Integrating over $k$ in the complex plane the third and
fourth factors of the integral vanish and we find that $\Gind{pp}$ is given
by
\begin{equation}
\Gind{pp} = L^2 {1\over \pi\sqrt{6\pi} \eta_0} {1\over\sqrt{L |p|}} \left[
{1\over\sqrt{l_A}} {L_A\over L} + {1\over\sqrt{l_B}}\left(1 - {L_A\over
L}\right)\right] \,\openone .
\end{equation}

The computation of the remaining elements $\Gind{pp'}|_{p\neq p'}$ and
$\Gind{0p}$ is accomplished in a similar manner and involves a rather lengthy
calculus. A simplification occurs if noticing that the $\Gind{pp'}$ operator
given by (\ref{app_B_Rouse}) is hermitian
($\Gind{pp'}=\Gind{p'p}^{\ast}$). Then we only need to calculate 
\begin{eqnarray}
\Gind{pp'} & = & {1\over 3\pi^2 \eta_0} \int_0^{\infty}\!dk\, \left\{ {L_A
	L\over i\pi (p'-p)}{(k \text{R}_A)^2 \over (k \text{R}_A)^4 + \left( {2
	\pi L_A\over L} \right)^2} \left[e^{{2 i\pi L_A\over L} (p' - p)} -
	1\right] \right.\\ 
	& & + \left. {L (L - L_A)\over i \pi (p'-p)} \left[ 1 - e^{{2 i \pi
	(L_A - L)\over L}(p'-p)}\right] {(k \text{R}_B)^2\over (k \text{R}_B)^4
	+ \left( {2 \pi (L-L_A)\over L} p \right)^2 }\right\} \,\openone
	,\nonumber 
\end{eqnarray}
which yields the final result
\begin{equation}
\Gind{pp'} = L^2 {1\over 2\pi^2 \sqrt{6\pi} \eta_0}{1\over \sqrt{L |p|}} \,
 	{1\over i (p'- p)} \left( {1\over \sqrt{l_B}} - {1\over \sqrt{l_A}}
 	\right)\left[1 - e^{{2 i \pi L_A\over L} (p'-p)}\right] \,\openone.
\end{equation}
Letting $p'\to 0$ gives $\Gind{p0}$ and from hermiticity we obtain $\Gind{0p} =
\Gind{p0}^{\ast}$. 

\section{Evaluation of the average
$\la\Sind{\lowercase{s}}\cdot\Sind{\lowercase{s'}} \ra$}
\label{app:SS}

We consider a Gaussian diblock copolymer A-B consisting in $N_A$ links of
Kuhn-length $l_A$ and $N_B$ links of Kuhn-length $l_B$, with $N_A + N_B =
N$. Let $\rvec_k$ be the bond vector (along the segment) of the $k$-th link.
Our goal is to construct the bi-variate probability distribution function
$P(\Sind{i},\Sind{j})$.\cite{YAM} This can be done by expressing the positions
$\Sind{i}$ and $\Sind{j}$ of the $i$-th and $j$-th beads about the
center-of-mass as
\begin{equation}
\Sind{i} = \sum_{k=1}^{N} \Psi_{ik} \rvec_k; \;\;\; \Sind{j} = \sum_{k=1}^{N}
\Psi_{jk} \rvec_k ,
\label{app:S_ij}
\end{equation}
where
\begin{equation}
\Psi_{ik} = \theta(i-k) + {k\over N+1} - 1; \; \Psi_{ik} = \theta(j-k) +
{k\over N+1} - 1 
\label{app:Psi}
\end{equation}
and $\theta(x)$ is the step-function. Also, we introduce the symmetric matrix
${\Bbb C}$
\begin{equation}
{\Bbb C} = \left(
	\begin{array}{cc}
	C_{ii} & C_{ij} \\
	C_{ji} & C_{jj}
	\end{array} \right) ,
\end{equation}
with its  components given by
\begin{equation}
C_{mn} = {1\over \la l^2 \ra} \left( l_A^2 \sum_{k=1}^{N_A} \Psi_{mk}\Psi_{nk}
+ l_B^2 \sum_{k=N_A+1}^{N} \Psi_{mk}\Psi_{nk} \right), \; m,n\in\{i,j\}
\label{app:C_components}
\end{equation}
and $\la l^2\ra = {1\over N}[N_A l_A^2 + (N - N_A) l_B^2 ] = l_A^2 (x + t^2 (1
- x))$.

With these notations, we can calculate in a well documented fashion \cite{YAM}
the bi-variate probability distribution function of the Gaussian diblock
copolymer A--B as
\begin{equation} 
P(\Sind{i},\Sind{j}) = \left( {3\over 2 \pi \la l^2 \ra}\right)^3 {1\over
(\det{\Bbb C})^{3/2}} \exp[- C_1 \Sind{i}\cdot\Sind{i} - C_2
\Sind{i}\cdot\Sind{j} -C_3 \Sind{j}\cdot\Sind{j}],
\label{app:probability}
\end{equation}
where 
\begin{equation}
C_1 = {3 C_{jj}\over 2 \la l^2\ra \det{\Bbb C}};\; C_2 = - {3 C_{ij}\over \la
l^2\ra  \det{\Bbb C}}; \; C_3 = {3 C_{ii}\over 2 \la l^2\ra \det{\Bbb C}}.
\end{equation}
The generating function for the moments is equal to:
\begin{equation}
M[\kvec,\kvec '] = \exp\left[ - {1\over 6}\la l^2 \ra (C_{ii} {\kvec}^2 + 2
C_{ij} \kvec\cdot\kvec ' + C_{jj} \kvec '^2 )\right]
\end{equation}
The sought average $\la \Sind{i}\cdot\Sind{j}\ra$ results immediately from:
\begin{equation}
\la \Sind{i}\cdot\Sind{j} \ra = \lim_{\stackrel{\scriptstyle \kvec\to 0}{\kvec
'\to 0}} i^2 \nabla_{\kvec}\nabla_{\kvec '}M[\kvec
,\kvec '] = \la l^2 \ra C_{ij}
\label{app:Si_Sj}
\end{equation}

It is clear that the value of the $C_{ij}$ factor depends on where the beads
$i$ and $j$ are located with respect to the two sections of the copolymer. We
have three cases: both beads in block A ($C_{ij}^A$), both beads in block B
($C_{ij}^B$) or the $i$-th bead in A and the $j$-th bead in B ($C_{ij}^{AB} =
C_{ji}^{BA}$). Inserting the values of $\Psi_{(i,j)k}$ from Eq.~(\ref{app:Psi})
in Eq.~(\ref{app:C_components}) and carrying out the summation keeping only
terms of order $N$, we obtain
\begin{mathletters}
\begin{eqnarray}
C_{ij}^A & = & {l_A^2\over \la l^2 \ra} \left\{{i^2\over 2 N} + {j^2\over 2 N}
- i \theta (i - j) - j [1 - \theta (i-j)] \right. \\ & & + \left. N_A \left(1 -
x + {x^2\over 3}\right) + t^2 {N\over 3}\left[1 - 3 x \left(1 - x + {x^2\over
3}\right) \right]\right\} \\ 
C_{ij}^B & = & {l_A^2\over \la l^2 \ra} \left\{
{N\over 3}[ t^2 + (1 - t^2) x^3 ] + t^2 \left[ {i^2\over 2 N} + {j^2\over 2 N}
- i \theta (i - j) - j [1 - \theta (i-j)] \right]\right\}\\ 
	C_{ij}^{AB} & = & {l_A^2\over \la l^2 \ra} \left\{ {i^2\over 2 N} -
{1\over 2} N_A x \left(1 - {2\over 3 }x \right) + t^2 \left[ {j^2\over 2 N} - j
+{N \over 3}\left(1 + {3\over 2 } x^2 - x^3 \right)\right]\right\},
\end{eqnarray}
\end{mathletters}
where we remind that $x = {N_A\over N}$ and $t = {l_B \over l_A}$. Returning to
the arclength variables $s$, $s'$ (e.g. $s = i l_A$ when $s \leq L_A$ or $s =
N_A l_A + (i - N_A) l_B$ for $s > N_A$), we complete the computation of
$\la\Sind{s}\cdot\Sind{s'}\ra$ by inserting the $C_{ij}$ factors in
(\ref{app:Si_Sj}). In the three mentioned cases of distribution of the beads
$s$ (former $i$) and $s'$ (former $j$) in the blocks A and B we find
\begin{mathletters}
\begin{eqnarray}
\la \Sind{s}\cdot\Sind{s'}\ra^A & = & {1\over 2 N} \left[s^2 + s'^2 - 2 s N l_A
\theta (s - s') - 2 s' N l_A (1 - \theta (s - s')) \right.\nonumber\\ 
& & + \left. 2 (N l_A)^2 \left(x (1 - t^2)(1 - x + {1\over 3} x^2) + {1\over 3}
t^2 \right) \right]\\
\la \Sind{s}\cdot\Sind{s'}\ra^B & = & {1\over 2 N} \left[ s^2 + s'^2 - 2 N l_A
[x (1 - t) - t \theta (s - s') ] s \right. \nonumber\\
& & - 2 N l_A [x (1 - t) + t (1 - \theta (s - s'))] s' \nonumber\\ 
& & + \left. 2 (N l_A)^2 \left({1 \over 3} x^2 (1 - t^2) + {t^2 \over 3} + x t
	(1 - t) + x^2 (1 - t)^2 \right) \right]\\
\la \Sind{s}\cdot\Sind{s'}\ra^{AB} & = & {1\over 2 N} \left[s^2 + s'^2 - 2 N
l_A (t + x(1 - t)) s' \right.\nonumber\\
& & + \left. 2 (N l_A)^2 \left( {1\over 2} x^2 (1 - t)^2 + x t(1 - t) + {t^2
\over 3} - {1\over 2} x^2 (1 - t^2) (1 - {2\over 3} x)\right)\right],
\end{eqnarray}
\end{mathletters}
with $\la \Sind{s}\cdot\Sind{s'}\ra^{BA} = \la
\Sind{s}\cdot\Sind{s'}\ra^{AB}|_{s \leftrightarrow s'}$.

Having calculated these averages, we introduce them back in
\begin{equation}
\subs{F}{pp'} = \text{F}_{pp'}\openone = \int\!\!\!\int_0^L\! \la\Sind{s}\cdot
\Sind{s'}\ra \, \exp[ -{2 i \pi \over L} p s + {2 i\pi\over L} p's'] \,
\openone .
\end{equation}
and after evaluating the resulting integrals we arrive at the Fourier
coefficients $\subs{F}{pp'}$. We show the $f_{00}$, $f_{pp}$ and
$f_{0p}$ elements (where $f_{pp'}$ is the dimensionless part of $\subs{F}{pp'}$
from Eq.~(\ref{F_dimless})):
\begin{mathletters}
\begin{eqnarray}
f_{00}(x,t) & = & {2\over 3} \, {{{( 1 - t) }^2} {{\left(1 - x \right) }^2}
{x^2} \left( {t^2} + x - {t^2} x \right) } \\
f_{p p}(x,t) & = & {1 \over {4\,{p^3}\,{{\pi }^3}}}
	{e^{{{-2\,i\,p\,\pi \,x}\over {t + x - t\,x}}}}\,{{\left( -t - x +
     t\,x \right) }^2}\, \left( -i\,t + i\,{e^{{{4\,i\,p\,\pi \,x}\over {t + x
     - t\,x}}}}\,t + i\,{t^2} - i\,{e^{{{4\,i\,p\,\pi \,x}\over {t + x -
     t\,x}}}}\,{t^2} \right.\\
	& & + 4\,{e^{{{2\,i\,p\,\pi \,x}\over {t + x -
     t\,x}}}}\,p\,\pi \,{t^2} - i\,x + i\,{e^{{{4\,i\,p\,\pi \,x}\over {t + x -
     t\,x}}}}\,x + 4\,{e^{{{2\,i\,p\,\pi \,x}\over {t + x - t\,x}}}}\,p\,\pi
     \,x  \nonumber\\
	& & + \left. 2\,i\,t\,x - 2\,i\,{e^{{{4\,i\,p\,\pi \,x}\over {t + x -
     t\,x}}}}\,t\,x - i\,{t^2}\,x + i\,{e^{{{4\,i\,p\,\pi \,x}\over {t + x -
     t\,x}}}}\,{t^2}\,x - 4\,{e^{{{2\,i\,p\,\pi \,x}\over {t + x -
     t\,x}}}}\,p\,\pi \,{t^2}\,x \right) \nonumber\\ 
f_{0p}(x,t) & = & {1 \over {4\,{{\pi }^3}\,{{{\it p}}^3}}}
	\left( -1 + t \right) \,\left( -t - x + t\,x \right) \,
     \left( i\,{t^2} - i\,{e^{{{2\,i\,\pi \,{\it p}\,x}\over {t + x -
     t\,x}}}}\,{t^2} \right. \\
	& & + 2\,i\,t\,x - 2\,i\,{e^{{{2\,i\,\pi \,{\it p}\,x}\over
     {t + x - t\,x}}}}\,t\,x - 2\,{e^{{{2\,i\,\pi \,{\it p}\,x}\over {t + x -
     t\,x}}}}\,\pi \,{\it p}\,t\,x - 2\,i\,{t^2}\,x + 2\,i\,{e^{{{2\,i\,\pi
     \,{\it p}\,x}\over {t + x - t\,x}}}}\,{t^2}\,x \nonumber\\ 
	& & + 2\,{e^{{{2\,i\,\pi
     \,{\it p}\,x}\over {t + x - t\,x}}}}\,\pi \,{\it p}\,{t^2}\,x +
     2\,i\,{{\pi }^2}\,{{{\it p}}^2}\,{t^2}\,x + i\,{x^2} - i\,{e^{{{2\,i\,\pi
     \,{\it p}\,x}\over {t + x - t\,x}}}}\,{x^2} \nonumber\\
	& & - 2\,{e^{{{2\,i\,\pi \,{\it
     p}\,x}\over {t + x - t\,x}}}}\,\pi \,{\it p}\,{x^2} + 2\,i\,{{\pi
     }^2}\,{{{\it p}}^2}\,{x^2} \nonumber\\
	& & - 2\,i\,t\,{x^2} + 2\,i\,{e^{{{2\,i\,\pi
     \,{\it p}\,x}\over {t + x - t\,x}}}}\,t\,{x^2} + 6\,{e^{{{2\,i\,\pi
     \,{\it p}\,x}\over {t + x - t\,x}}}}\,\pi \,{\it p}\,t\,{x^2} +
     i\,{t^2}\,{x^2} - i\,{e^{{{2\,i\,\pi \,{\it p}\,x}\over {t + x -
     t\,x}}}}\,{t^2}\,{x^2} \nonumber\\
	& & - 4\,{e^{{{2\,i\,\pi \,{\it p}\,x}\over {t + x -
     t\,x}}}}\,\pi \,{\it p}\,{t^2}\,{x^2} - 4\,i\,{{\pi }^2}\,{{{\it
     p}}^2}\,{t^2}\,{x^2} + 2\,{e^{{{2\,i\,\pi \,{\it p}\,x}\over {t + x -
     t\,x}}}}\,\pi \,{\it p}\,{x^3} - 2\,i\,{{\pi }^2}\,{{{\it
	p}}^2}\,{x^3}\nonumber\\ 
     & & - \left. 4\,{e^{{{2\,i\,\pi \,{\it p}\,x}\over {t + x - t\,x}}}}\,\pi
     \,{\it p}\,t\,{x^3} + 2\,{e^{{{2\,i\,\pi \,{\it p}\,x}\over {t + x -
     t\,x}}}}\,\pi \,{\it p}\,{t^2}\,{x^3} + 2\,i\,{{\pi }^2}\,{{{\it
     p}}^2}\,{t^2}\,{x^3} \right)\nonumber
\end{eqnarray}
\end{mathletters}
Unfortunately, $f_{pp'}(x,t)$ is too long to be reproduced here.  Limits of
$f_{pp'}(x,t)$ when $x = \{0,1\}$ and $t = 1$ were quoted in the text (see
Eq.\ (\ref{f_limits})).

\begin{figure}
\begin{center}
	\begin{tabular}{c}
		\begin{minipage}{17 true cm}
\vspace*{-0.5cm}
$${\epsfxsize=11.2 true cm \epsfbox{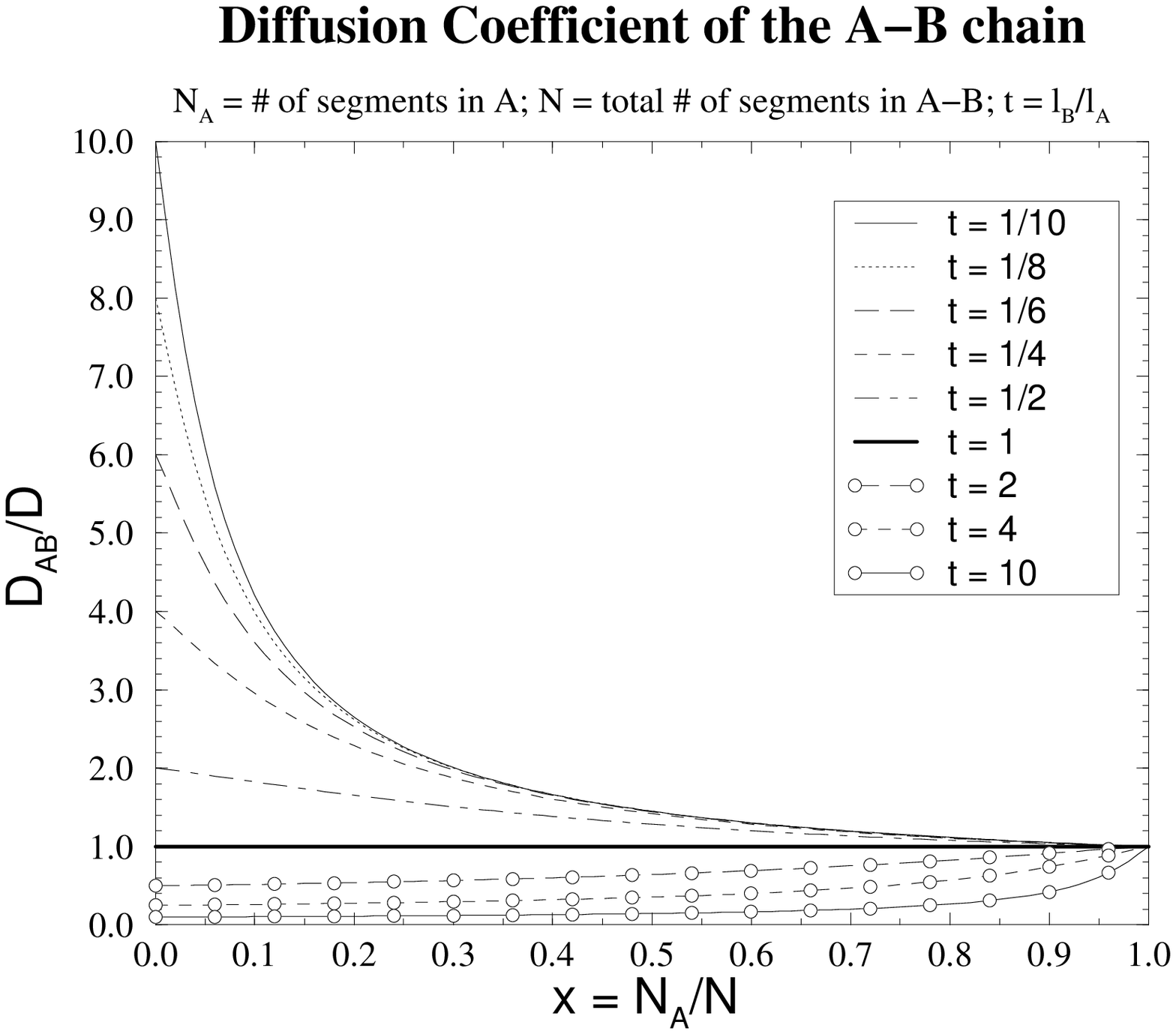}}$$
\caption{Diffusion coefficient $D_{\text{AB}}/D$ as a function of $x = N_A/N$
and $t = l_B/l_A$ ($D$ is the diffusion coefficient of a homogeneous chain with
length $L_N = N l_A$) : {\sf solid line}\ $t = 1/10$; {\sf dotted line}\ $t =
1/8$; {\sf long dashed line}\ $t = 1/6$; {\sf dashed line}\ $t = 1/4$; {\sf
dot-dashed line}\ $t = 1/2$; {\sf thick solid line}\ $t = 1$; $\circ$-{\sf
long dashed line}-$\circ$~$t = 2$; $\circ$-{\sf dashed
line}-$\circ$~$t = 4$; $\circ$-{\sf solid
line}-$\circ$~$t = 10$ .} \label{fig:diffAB}
$${\epsfxsize=11.2 true cm \epsfbox{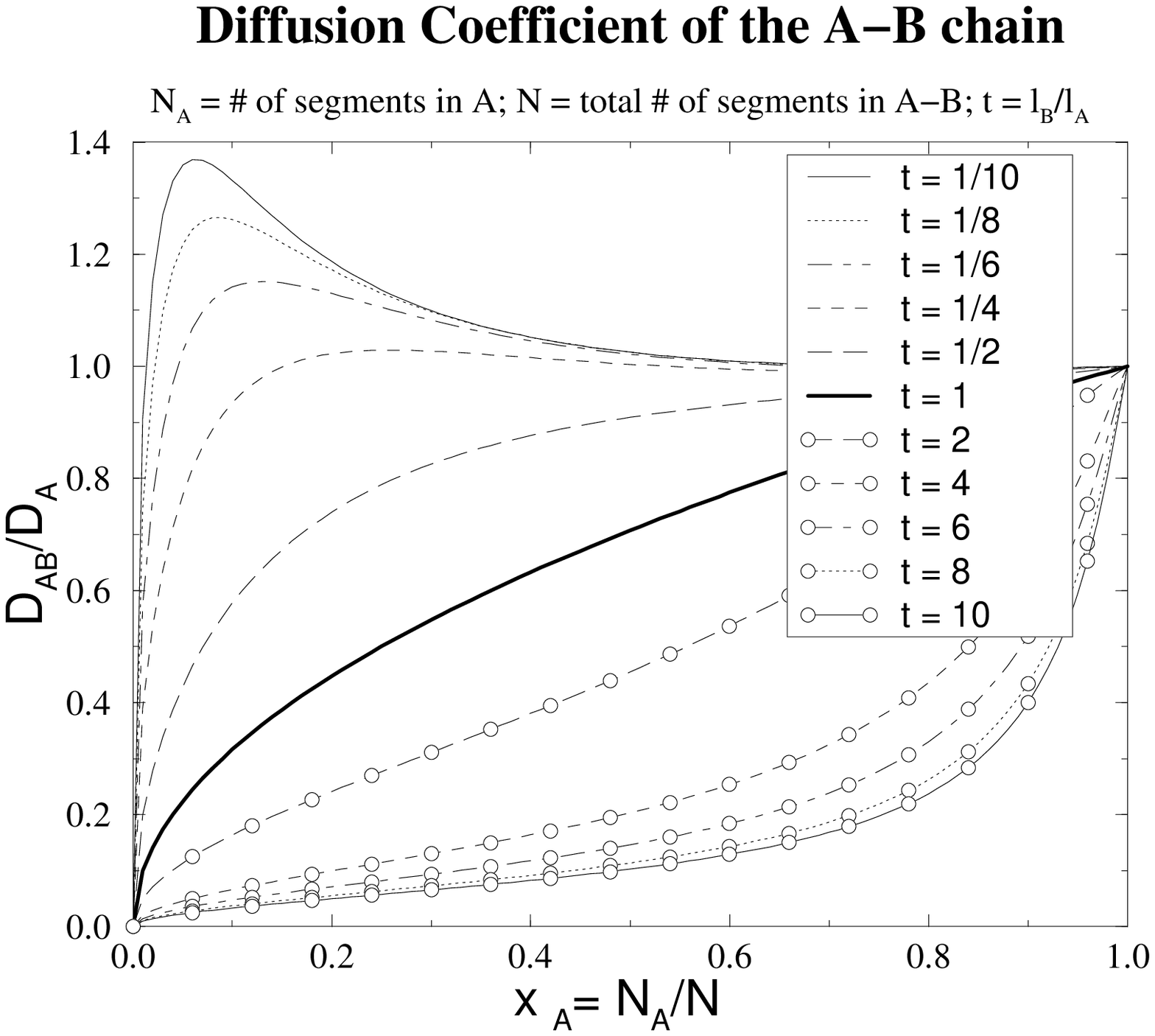}}$$
\caption{Diffusion coefficient $D_{\text{AB}}/D_A$ as a function of $x = N_A/N$
and $t = l_B/l_A$ ($D_A$ is the diffusion coefficient of a homogeneous chain
with length $L_A = N_A l_A$): {\sf solid line}\ $t = 1/10$; {\sf dotted line}\
$t = 1/8$; {\sf dot-dashed line}\ $t = 1/6$; {\sf dashed line}\ $t = 1/4$; {\sf
long dashed line}\ $t = 1/2$; {\sf thick solid line}\ $t = 1$; $\circ$-{\sf
long dashed line}-$\circ$~$t = 2$; $\circ$-{\sf dashed
line}-$\circ$~$t = 4$; $\circ$-{\sf dot-dashed line}-$\circ$~$t = 6$;
$\circ$-{\sf dotted line}-$\circ$~$t = 8$; $\circ$-{\sf solid
line}-$\circ$~$t = 10$ .} \label{fig:diffABA}
		\end{minipage}
	\end{tabular}
\end{center}
\end{figure}

\begin{figure}
\begin{center}
	\begin{tabular}{c}
		\begin{minipage}{17cm}
\vspace*{-0.5cm}
$${\epsfxsize=11.5 true cm \epsfbox{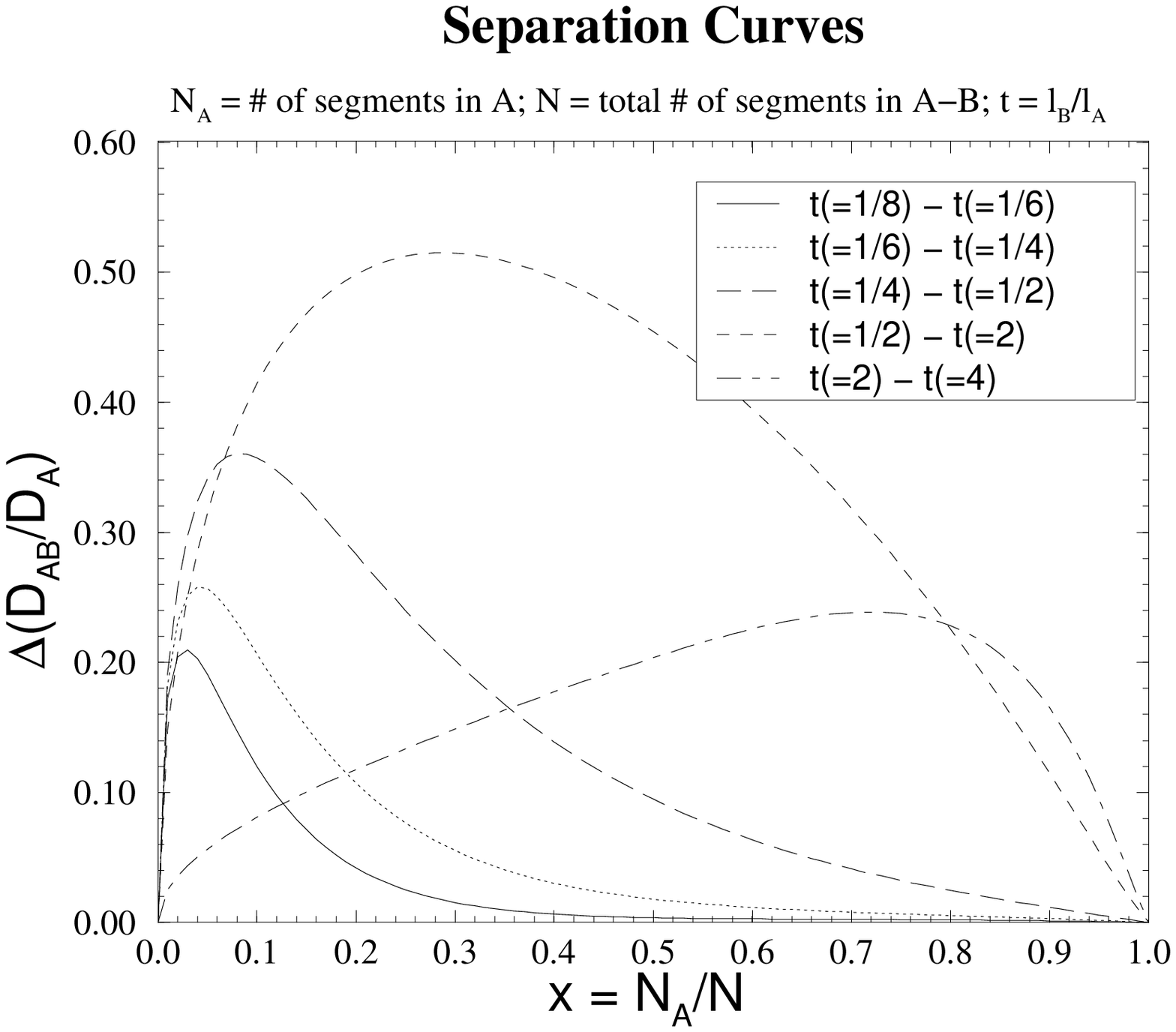}}$$
\caption{Separation curves as $\Delta(D_{\text{AB}}/D_A)(x)={D_{\text{AB}}\over
D_A}(x,t_1)-{ D_{\text{AB}}\over D_A}(x,t_2)$:  {\sf solid
line}\ $[t_1 = 1/8;t_2 = 1/6]$; {\sf dotted line}\ $[t_1 = 1/6;t_2 = 1/4]$;
{\sf long dashed line}\ $[t_1 = 1/4;t_2 = 1/2]$; {\sf dashed line}\ $[t_1 =
1/2;t_2 = 2]$; {\sf dot-dashed line}\ $[t_1 = 2;t_2 = 4]$ .}
			\label{fig:diffSEP}
$${\epsfxsize=11.5 true cm \epsfbox{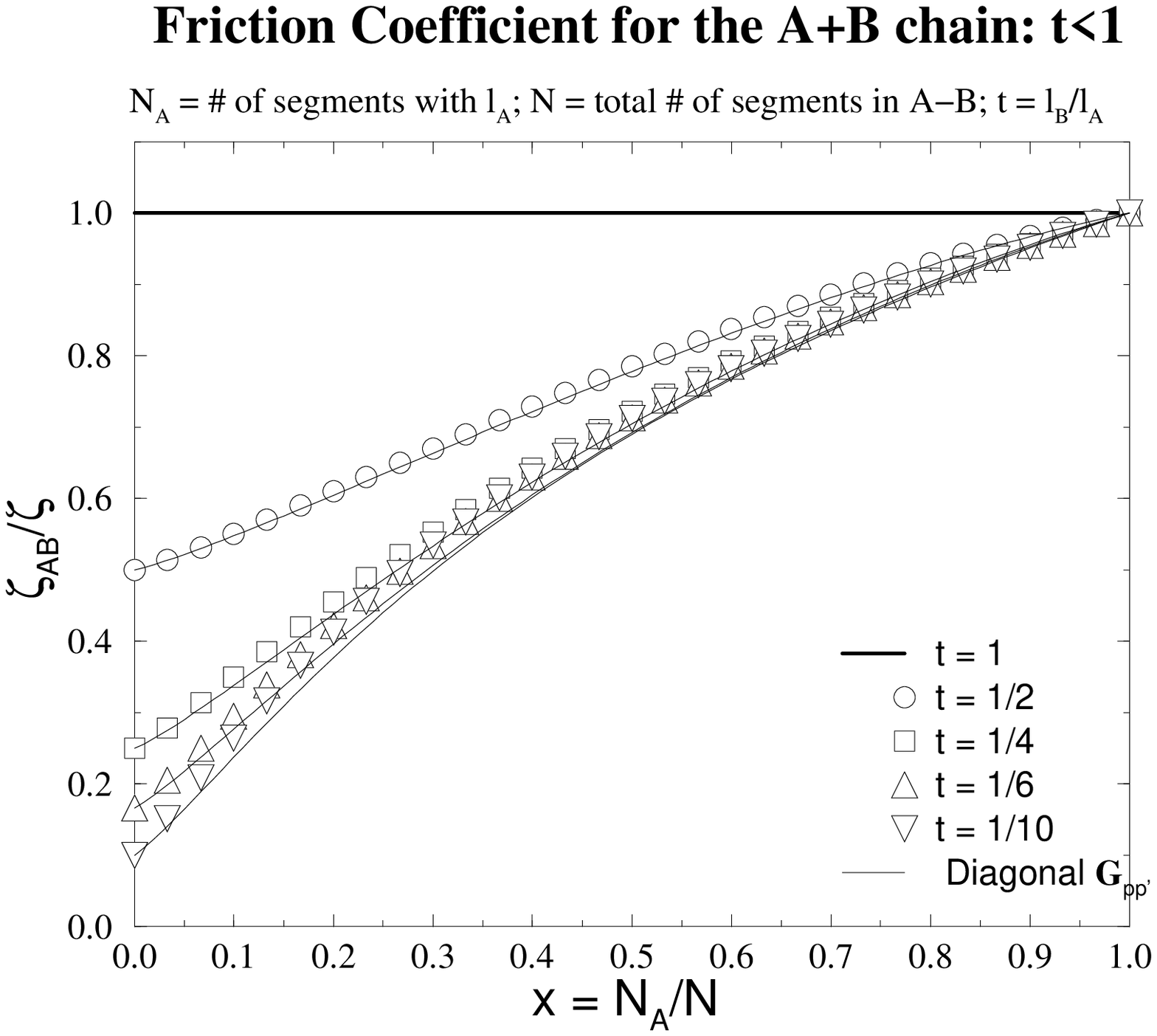}}$$
\caption{Friction coefficient $\zeta_{\text{AB}}/\zeta$ as a function of $x$
and $t$ ($\zeta$ is the Kirkwood-Riseman (K-R) friction coefficient for a chain
with length $L_N = N l_A$). Thin lines show the results obtained when using
only the diagonal elements of $\Gind{pp'}$ for the $t$ values corresponding to
the symbols: {\sf thick solid line}\ $t = 1$; $\circ$\ $t = 1/2$; $\square$\ $t
= 1/4$; $\bigtriangleup$\ $t = 1/6$; $\bigtriangledown$\ $t = 1/10$; ---\
Diagonal $\Gind{pp'}$.}  \label{fig:frLess} \end{minipage} \end{tabular}
\end{center}
\end{figure}

\begin{figure}
\begin{center}
	\begin{tabular}{c}
		\begin{minipage}{17cm}
\vspace*{-1cm}
$${\epsfxsize=11.5 true cm  \epsfbox{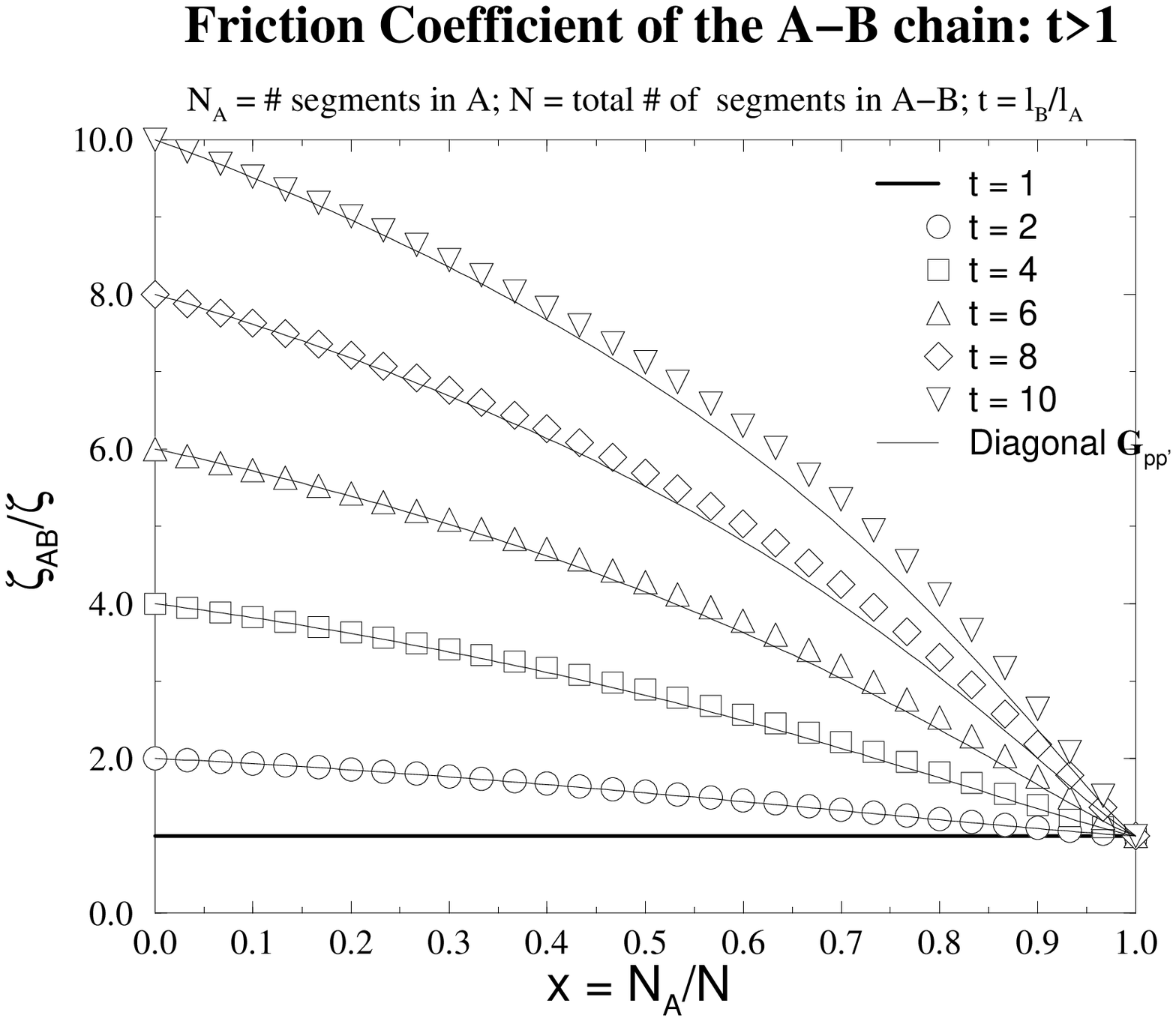}}$$
\caption{Friction coefficient $\zeta_{\text{AB}}/\zeta$ as a function of $x$
and $t$ ($\zeta$ is the K-R friction coefficient for a chain with length $L_N =
N l_A$ ). Thin lines show the results obtained when using only the diagonal
elements of $\Gind{pp'}$ for the $t$ values corresponding to the symbols: {\sf
thick solid line}\ $t = 1$; $\circ$\ $t = 2$; $\square$\ $t = 4$;
$\bigtriangleup$\ $t = 6$; $\Diamond$\ $t = 8$; $\bigtriangledown$\ $t = 10$;
---\ Diagonal $\Gind{pp'}$ .} 
\label{fig:frBig}
$${\epsfxsize=12 true cm \epsfbox{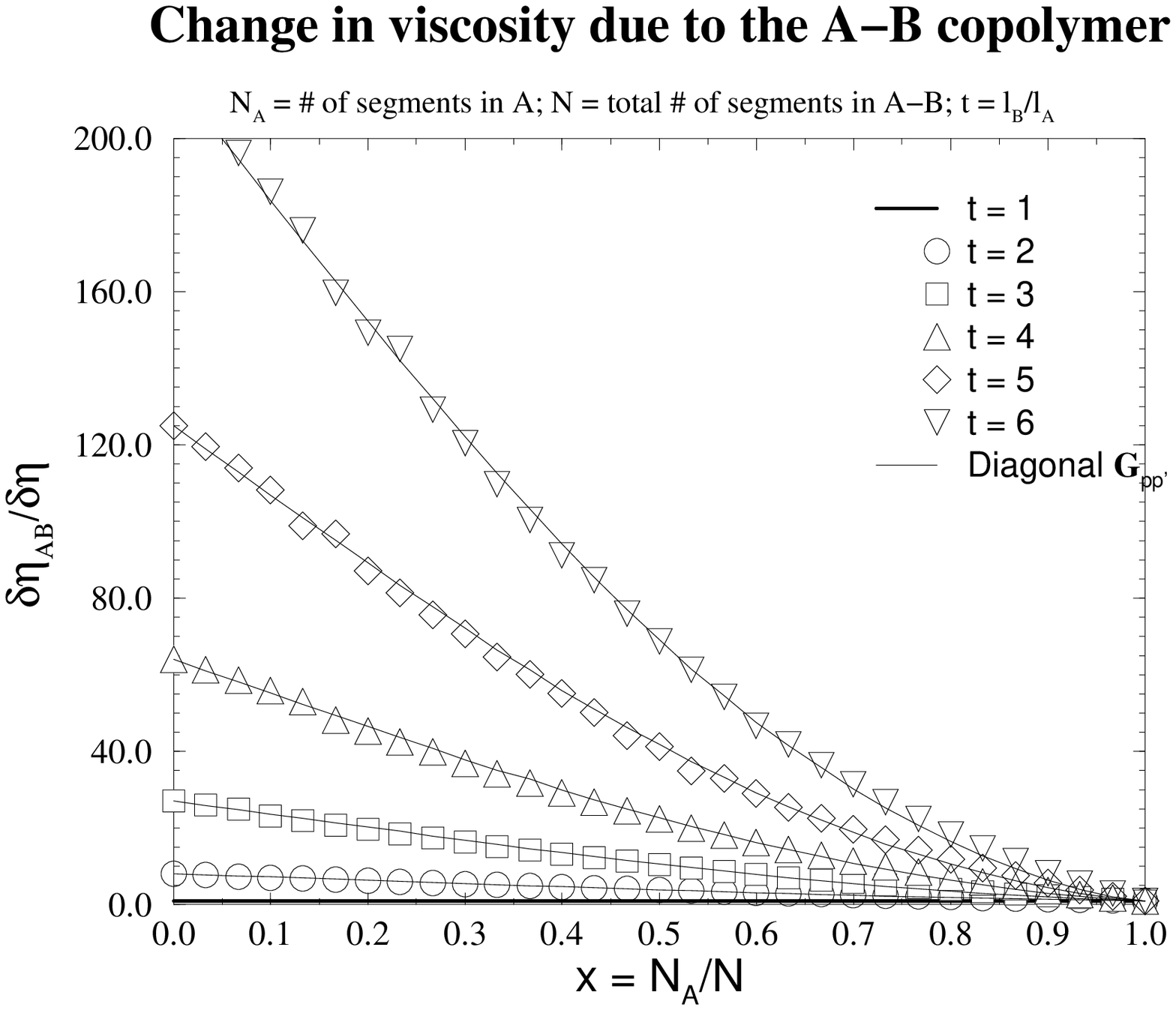}}$$
\caption{Change in the viscosity $\delta\eta_{\text{AB}}/\delta\eta$ as a
function of $x$ and $t$ ($\delta\eta$ is the K-R non-free draining result for
the change in viscosity due to a polymer with length $L=N l_A$). Thin lines
show the results obtained when using only the diagonal elements of $\Gind{pp'}$
for the $t$ values corresponding to the symbols: {\sf thick solid line}\ $t =
1$; $\circ$\ $t = 2$; $\square$\ $t = 3$; $\bigtriangleup$\ $t = 4$;
$\Diamond$\ $t = 5$; $\bigtriangledown$\ $t = 6$; ---\ Diagonal $\Gind{pp'}$ .}
\label{fig:viscosity}
		\end{minipage}
	\end{tabular}
\end{center}
\end{figure}

\begin{figure}
\begin{center}
	\begin{minipage}{17cm}
$${\epsfxsize=14 true cm \epsfbox{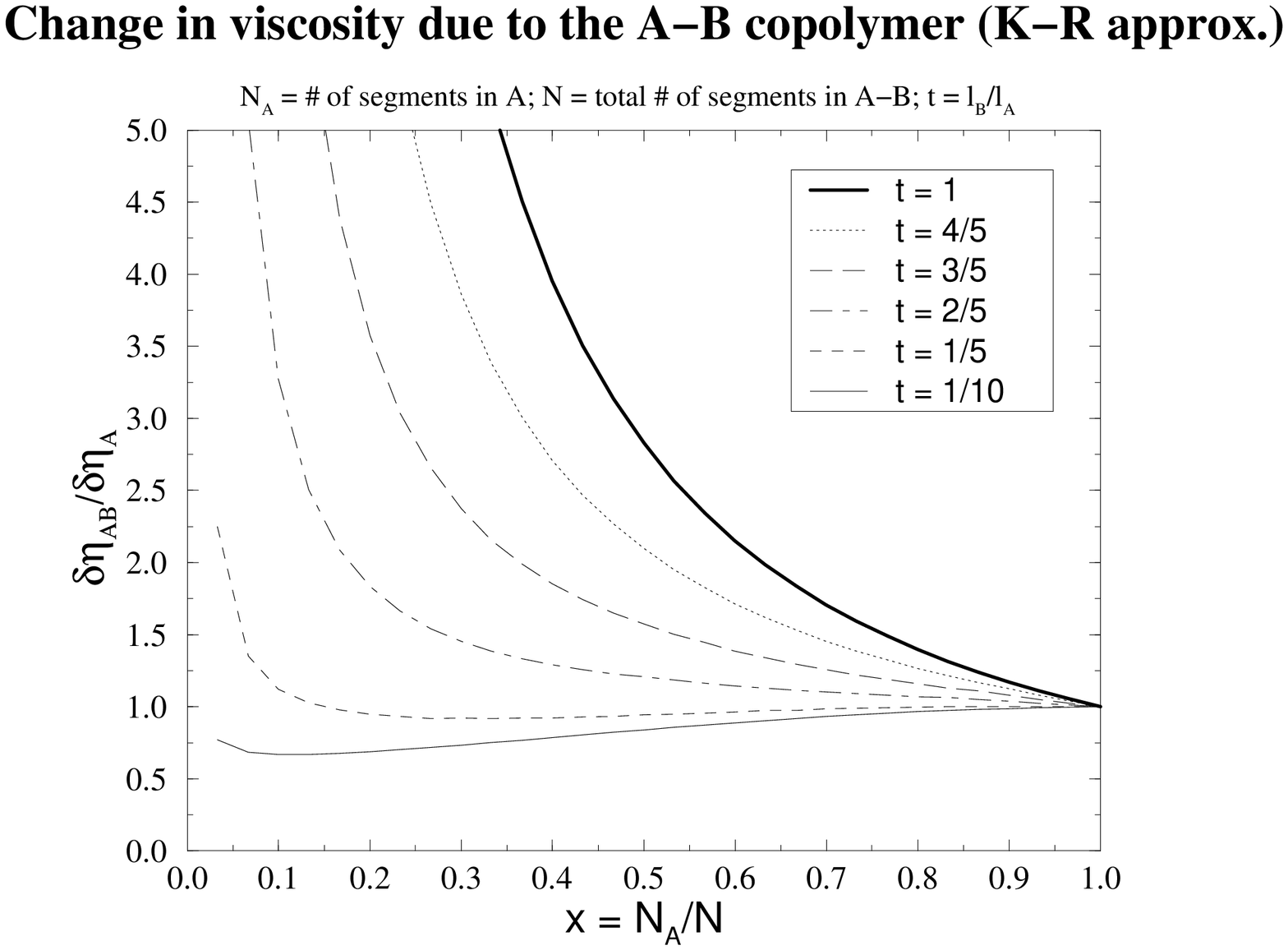}}$$
\caption{Change in the viscosity $\delta\eta_{\text{AB}}/\delta\eta_A$ as a
function of $x$ and $t$ using the Kirkwood-Riseman approximation
($\delta\eta_A$ is the K-R non-free draining result for the change in viscosity
due to a polymer with length $L_A=N_A l_A$): {\sf thick solid line}\ $t = 1$;
{\sf dotted line}\ $t = 4/5$; {\sf long dashed line}\ $t = 3/5$; {\sf
dot-dashed line}\ $t = 2/5$; {\sf dashed line}\ $t = 1/5$; {\sf solid line}\ $t
= 1/10$ .}
\label{fig:viscosityABA}
	\end{minipage}
\end{center}
\end{figure}

\end{document}